%% file: Main.tex
 \documentclass[final,5p,times,twocolumn,authoryear]{elsarticle}

\usepackage{amssymb}
\usepackage{amsmath}
\usepackage{mathtools}
\usepackage{bm}
\usepackage{units}
\usepackage{subcaption}
\usepackage{enumitem}
\usepackage{etoolbox}
\usepackage{accents}
\usepackage{threeparttable}
\usepackage{tabularx}
\newcolumntype{Y}{>{\centering\arraybackslash}X}

\RequirePackage{booktabs,makecell,multirow,array,colortbl,dcolumn,stfloats}
\usepackage[dvipsnames]{xcolor}

\RequirePackage[colorlinks]{hyperref}
\colorlet{scolor}{black}
\colorlet{hscolor}{black}
\hypersetup{%
  linkcolor={hscolor},
  urlcolor={hscolor},
  citecolor={hscolor},
  filecolor={hscolor},
  menucolor={hscolor},
}

\usepackage{nomencl}

\renewcommand\nomgroup[1]{%

  \ifstrequal{#1}{V}{\medskip \item[\textit{Select Variables}]}{%
  \ifstrequal{#1}{B}{\medskip \item[ \textit{Subscripts}]}{%
  \ifstrequal{#1}{P}{\medskip \item[ \textit{Notation}]}{%
  \ifstrequal{#1}{A}{\item[ \textit{Acronyms}]}{}}}}
}

\renewcommand*\nompreamble{\begin{multicols}{2}}
\renewcommand*\nompostamble{\end{multicols}}

\makenomenclature

\usepackage{algorithm2e}
\usepackage{xcolor}

\SetAlFnt{\footnotesize}

\SetAlCapSty{xAlCapSty}

\definecolor{commentcolor}{HTML}{1E4D2B}

\SetCommentSty{xCommentSty}

\SetNlSty{mynlfont}{}{} 

\LinesNumbered

\SetSideCommentRight

\DontPrintSemicolon

\RestyleAlgo{algoruled}

\usepackage{algorithm2e}
\usepackage{etoolbox}
\usepackage{xstring}
\usepackage{calc}
 
\newlength{\xalgowidth}
\newlength{\xalgoremainder}
\newlength{\xindentwidth}

\newenvironment{vAlgorithm*}[3][]{
  \setlength{\xalgowidth}{#2} 
  \setlength{\xindentwidth}{#3} 
  \setlength{\xalgoremainder}{\textwidth-\xalgowidth} 
  \SetCustomAlgoRuledWidth{\xalgowidth} 
  \IncMargin{\xindentwidth}
  \begin{algorithm*}[#1]
}
{
  \end{algorithm*} 
  \DecMargin{\xindentwidth}
}

\newenvironment{vAlgorithm}[3][]{
  \setlength{\xalgowidth}{#2} 
  \setlength{\xindentwidth}{#3} 
  \setlength{\xalgoremainder}{\columnwidth-\xalgowidth} 
  \SetCustomAlgoRuledWidth{\xalgowidth} 
  \IncMargin{\xindentwidth}
  \begin{algorithm}[#1] 
}
{
  \end{algorithm} 
  \DecMargin{\xindentwidth}
}

\makeatletter
\patchcmd{\@algocf@start}{%
\begin{lrbox}{\algocf@algobox}%
}{%
\rule{0.5\xalgoremainder}{\z@}
\begin{lrbox}{\algocf@algobox}%
\begin{minipage}{\xalgowidth}%
}{}{}
\patchcmd{\@algocf@finish}{%
\end{lrbox}%
}{%
\end{minipage}%
\end{lrbox}%
}{}{}
\makeatother

\newcommand{\MyBold}[1]{\mathbf{#1}}
\newcommand{\parm}{\mathord{\color{black!33}\bullet}}%

\newcommand{\Rwec}{R_{\text{wec}}}            
\newcommand{\Dwec}{D_{\text{wec}}}            
\newcommand{\RDwec}{RD_{\text{wec}}}          
\newcommand{\Rwecmax}{\bar{R}_{\text{wec}}}   
\newcommand{\Dwecmax}{\bar{D}_{\text{wec}}}   
\newcommand{\RDwecmax}{\bar{RD}_{\text{wec}}}   
\newcommand{\Rwecmin}{\underaccent{\bar}{R}_{\text{wec}}}   
\newcommand{\Dwecmin}{\underaccent{\bar}{D}_{\text{wec}}}   
\newcommand{\RDwecmin}{\underaccent{\bar}{RD}_{\text{wec}}}   

\newcommand{\Angpq}{\theta_{pq}}              
\newcommand{\Dispq}{l_{pq}}                   

\newcommand{\Nwec}{n_{\text{wec}}}     
\newcommand{\NMBE}{m_{\text{}}}      
\newcommand{\Mass}{\MyBold{M}}       
\newcommand{\Force}{\MyBold{F}}      
\newcommand{\AddedMass}{\MyBold{A}}     
\newcommand{\DampingCoeff}{\MyBold{B}}  
\newcommand{\addedmassv}{{a}}       
\newcommand{\dampingcoeffv}{{b}}    

\newcommand{\OMF}{(\omega)}             
\newcommand{\AL}{\MyBold{w}}           
\newcommand{\KPTO}{\MyBold{K}_{\text{pto}}}           
\newcommand{\BPTO}{\MyBold{B}_{\text{pto}}}           
\newcommand{\KPTOmin}{\underaccent{\bar}{\MyBold{k}}_{\text{pto}}}           
\newcommand{\BPTOmin}{\underaccent{\bar}{\MyBold{B}}_{\text{pto}}}           
\newcommand{\KPTOmax}{\bar{\MyBold{k}}_{\text{pto}}}           
\newcommand{\BPTOmax}{\bar{\MyBold{B}}_{\text{pto}}}           

\newcolumntype{s}{>{\columncolor[HTML]{F5F5F5}} c}

\makeatletter
\newcommand*{\rom}[1]{\expandafter\@slowromancap\romannumeral #1@}
\makeatother

\usepackage{mdframed} 

\usepackage{multicol} 

\usepackage{printlen} 

\newcommand{\xywecscale}{0.75}
\newcommand{\twofigscale}{0.52}

\journal{Renewable Energy}

\begin{document}

\input{Front}


\section{Introduction}\label{sec:introduction}

With its unique characteristics, such as temporal and spatial availability, low variability, and high predictability, wave energy is recognized as a promising source of renewable energy \citep{ning2022modelling}.
Despite significant advancements since their conception in 1799 \citep{Ross2012}, commercial wave energy converter (WEC) devices still face challenges.
Particularly, their technology readiness level (TRL), which is often used to assess the development maturity of a new technology, is still low compared to wind and solar technologies \citep{straub2015search}.
In addition to the hostile and corrosive ocean environment and inherent uncertainties in the wave climate, there is a lack of agreement regarding the most effective concept for harnessing wave energy \citep{ringwood2023empowering}.  
Therefore, further research is required to hasten the development of more cost-effective and resilient WEC devices.

\begin{table*}[!t]
\begin{mdframed}
\input{Nomenclature}
\vspace{-2\baselineskip}
\setlength{\nomitemsep}{-\parsep}
\printnomenclature
\end{mdframed}
\end{table*}

The wide technological diversity of WECs, along with the lack of convergence on the technology type and operating principles, motivate not only the development of suitable metrics to quantify technology development status \citep{weber2012wec}, but also the full realization of emerging concepts in design theory, such as control co-design (CCD) \citep{GarciaSanz2019}, which offers a promising solution to improve technology performance level by leveraging domain coupling.
This domain coupling refers to the fact that in many complex dynamic systems (including WECs), the \textit{optimized} plant (such as WEC geometry) changes when control methods or parameters (such as power take-off parameters) are changed (and vice versa\textemdash therefore the coupling is bidirectional). 
CCD concurrently solves a plant and control optimization problem and, consequently, leverages the coupling between various system domains to find a superior, system-level optimal solution. 
Therefore, any effort towards creating consensus within the community for convergence on a single WEC technology is better supported if the underlying performance evaluation method utilizes CCD.

To improve the economic viability of WECs, they must be carefully deployed in a farm layout. 
While reducing the installation, maintenance, and operational costs \citep{abdulkadir2023optimization}, the presence of multiple WECs in close proximity results in a complex hydrodynamic interaction effect \citep{falnes1980radiation}.
Since this effect can be constructive or destructive \citep{borgarino2012impact}, the spatial configuration of WECs within an array is strategically determined through array or layout optimization \citep{abdulkadir2023optimization, neshat2022layout, mercade2017layout}.
Since WEC geometry and control directly affect the optimal configuration of the WECs (and vice versa), these domains are also coupled \citep{Azad2024}.
Due to their time-independence and generally fixed determination during initial deployment, geometry and layout variables can be classified as \textit{plant} decision variables in CCD frameworks.
Leveraging this domain coupling through CCD in the design of WEC devices leads to superior system-level solutions that are not achievable through sequential design approaches \citep{GarciaSanz2019, ringwood2023empowering, Azad2023}. 
The coupling between WEC geometry, control, and layout is further discussed and investigated in \cite{Azad2024}.

With a few exceptions, the primary WEC literature has focused on the optimization of individual design domains such as physical WEC geometry (e.g.,~radius, slenderness ratio, draft) \citep{GarciaTeruel2021, Guo2021}, power take-off (PTO) control parameters \citep{RingwoodEnergyMax, Wang2018}, or layout (position of WEC devices within a farm) \citep{neshat2022layout, mercade2017layout}.
Largely based on a sequential design paradigm, these studies overlook domain coupling that can be used to improve the performance of WEC farms. 
Some recent studies, however, included domain integration in the design of WECs.
While the bulk of such studies focus on leveraging the coupling between geometry and control of a single WEC device, \citep{herber2013wave, strofer2023control, pena2022control, coe2020initial, Abdulkadir2024}, a few have integrated the layout optimization problem concurrently with geometry or control optimization \citep{GarciaRosa2015, lyu2019optimization, abdulkadir2023optimization, PenaSanchez2024}.

In addition to WEC device specifications and configuration, the performance of the WEC farm is directly influenced by its location (i.e., site location).
The site location, generally characterized by the naturally-available wave energy resource \citep{Jacobson2011, OceanEnergy}, dictates crucial environmental factors such as water depth, wave direction, significant wave height, wave period, and their joint probability distribution.
All of these factors impact the optimal WEC farm design, which in this article refers to the determination of the physical decisions (i.e., geometric attributes), control parameters, and layout configuration of the WEC devices within a farm.
While several studies in the WEC farm literature consider site selection \citep{tan2021influence,neshat2022layout,de2016techno,o2013techno}, they are largely limited to single-domain investigations.  

\begin{figure*}
    \centering
    \includegraphics[width=0.9\textwidth]{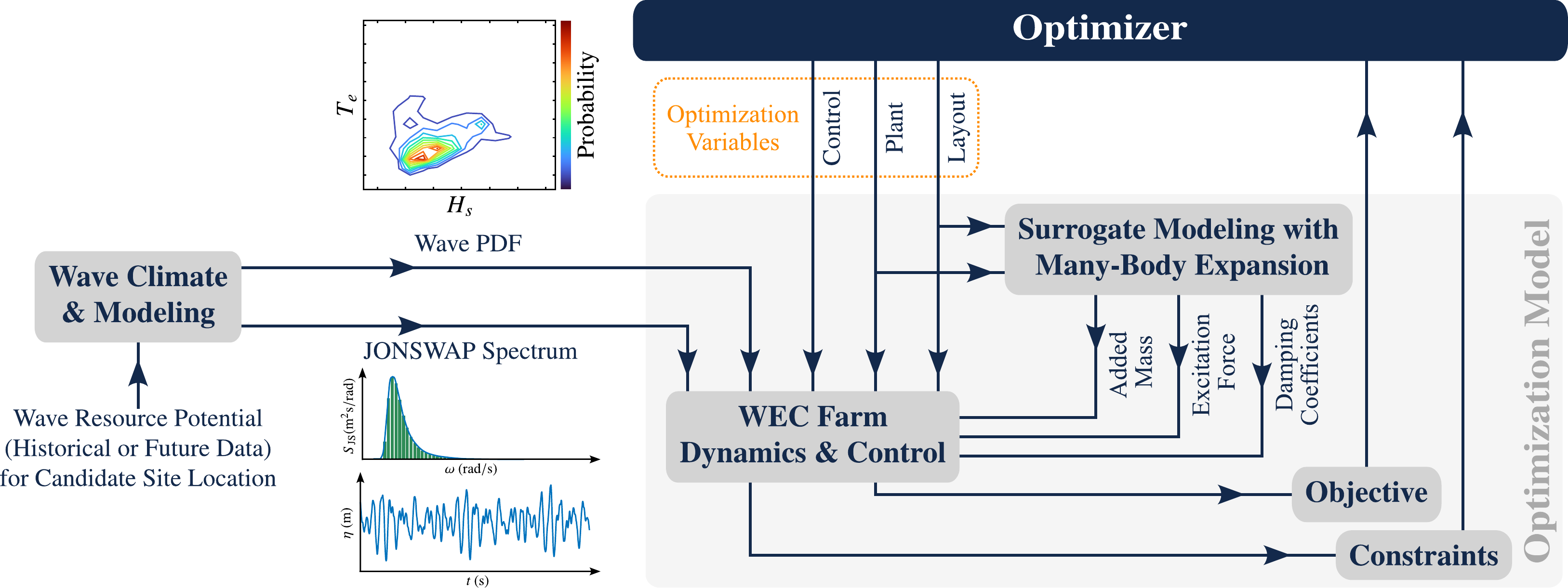}
    \caption{The general framework for the concurrent plant, control, and layout optimization using surrogate modeling and principles of MBE. The joint probability distribution of significant wave heights and wave periods for various US locations are constructed based on historical data.}
    \label{fig:Wave_Sys}
\end{figure*}

By leveraging a system-level design framework that concurrently considers the entire device and its characteristic attributes, such as plant, control, and layout, potential improvements in WEC array performance may be realized \citep{ringwood2023empowering}.
However, solving the resulting extensive and complex optimization problem is hindered by the need for accurate estimation of the hydrodynamic coefficients \citep{lyu2019optimization}. 
As an example, depending on the body mesh and computer specifications, a single call\footnote{The computational times in this article are reported using a desktop workstation with an AMD Ryzen Threadripper 3970X 12-core processor at 3.79 \unit{GHz}, 32 \unit{GB} of RAM, 64-bit windows 10 Enterprise LTSC version 1809, and \texttt{Matlab} R2022a} to the boundary element method (BEM) solver \texttt{Nemoh~version v113} \citep{babarit2015theoretical, Kurnia2022, kurnia2022second} for a two- and three-WEC farm can take about $384~[\unit{s}]$, and $1305~[\unit{s}]$, respectively.
Multiple scattering (MS) techniques, while offering a more efficient iterative scheme, still entail some computationally costly steps \citep{zhang2020surrogate}, taking $2.23~[\unit{s}]$ and $2.69~[\unit{s}]$, respectively. 
This computational cost, which increases dramatically for farms with a higher number of WECs, has hindered more complex integrated WEC design studies in the literature, resulting in investigations of fixed array configurations such as square and triangular \citep{borgarino2012impact}, and simplified assumptions such as regular waves \citep{lyu2019optimization}.

In this article, the computational bottleneck associated with the estimation of hydrodynamic coefficients is mitigated by constructing data-driven surrogate models (SMs) using artificial neural networks (ANNs), along with a hierarchical interaction decomposition approach inspired by concepts from many-body expansion (MBE) \citep{gordon2012fragmentation, zhang2020surrogate}.
Theoretically, the training data for estimating the hydrodynamic interaction effect can be developed through appropriate techniques, such as BEM or MS.
Due to superior computational efficiency and numerical smoothness when compared to the BEM software \texttt{Nemoh}, MS was used to develop data for the training of the SMs.
Note that the numerical smoothness of \texttt{Nemoh}'s solution was challenged by the presence of irregular frequencies \citep{kelly2022post}, an issue that has been resolved in recent releases of the software \citep{nemoh, BEMRosetta}.

The resulting SMs are then used inside the concurrent optimization framework to implement a surrogate-assisted optimization approach, \textit{efficiently} addressing the issue of accurate estimation of hydrodynamic coefficients within large arrays, and through that, enabling integrated design investigations with increasing complexity. 
This framework, implemented in the presence of probabilistic irregular waves, naturally formulates a stochastic in expectation uncertain control co-design (SE-UCCD) \citep{Azad2023} problem.
By enabling the simultaneous implementation of CCD \citep{GarciaSanz2019, strofer2023control, coe2020initial, o2017co, pena2022control, Azad2023a} with layout optimization for half-submerged cylindrical, heaving WEC devices in the presence of uncertainties from incident waves, this article paves the way for more complex and practical design investigations of WEC farms with more realistic assumptions.

Figure \ref{fig:Wave_Sys} shows the general framework for the concurrent plant, control, and layout optimization using SMs with the principles of MBE.
In this article, we model the heaving WEC devices in the frequency domain, with a PTO system that exerts a load force on the oscillating body while storing energy.
This PTO provides a bi-directional power flow through a reactive control strategy \citep{ning2022modelling}.
The absorbed power by the WEC device is calculated using the probability distribution of waves, constructed based on $30$ years of historical data collected from four representative site locations in Alaska Coasts, East Coast, Pacific Islands, and West Coast.
Defining the average power per unit volume of the wave-absorbing body as the optimization objective \citep{falnes2020ocean}, various case studies with increasing levels of complexity are investigated.
The open-source toolbox resulting from this research effort is publicly accessible in \cite{code}.

The remainder of this article is organized as follows:
Sec.~\ref{sec:section2} presents the methodological discussion, including probabilistic wave modeling, wave-structure interactions, dynamics and equations of motions for WECs, array geometry and considerations, SM with an emphasis on ANN, and a brief introduction into principles of MBE.
Sec.~\ref{sec:Constructing_Surrogate_Models} describes the construction of SMs using ANNs and principles of MBE through an active learning strategy and provides several methods for their validation through simulation and optimization studies.
Sec.~\ref{sec:Optimization} starts with discussing some of the limitations and challenges of a system-level optimization when using frequency-domain models and then presents results and discussion for various case studies with increasing levels of complexity including: plant optimization; concurrent plant and layout optimization; plant, layout, and farm-level control optimization; and plant, layout, and device-level control optimization. 
This section is accompanied by further CCD investigations with an increasing number of WEC devices within the farm.
Finally, Sec.~\ref{sec:conclusion} presents conclusions and limitations of the current study, as well as potential future work.

\section{Background}\label{sec:section2}

In this section, we describe the wave climate and modeling for specific site locations, followed by a brief description of wave-structure interactions.
The dynamics and control of WECs are then presented, and array considerations are discussed.  
This section concludes with a brief introduction of surrogate modeling for hydrodynamic interactions among WEC devices. 

\subsection{Wave Climate and Modeling} \label{subsec:WaveClimate}

\begin{figure}
    \centering
    \includegraphics[width=\columnwidth]{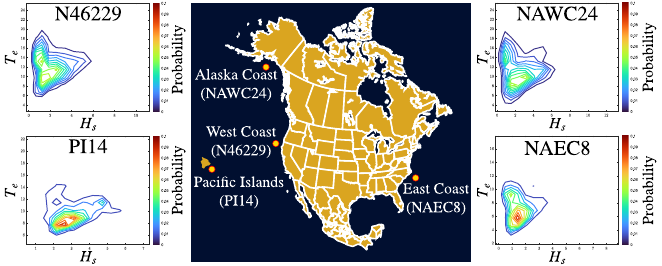}
    \caption{Joint distribution of $H_s$ and $T_p$ for the first year of the study at the selected site locations on the Alaskan Coast, East Coast, Pacific Islands, and West Coast based on data from~\cite{storlazzi2015future, erikson2016wave}.}
    \label{fig:Wave_loc}
\end{figure}

Wave energy applications consider waves resulting from the wind blowing over the ocean surface, dominated by gravity and inertial forces \citep{ning2022modelling}.
These waves are known as wind-generated gravity surface waves.
Four representative site locations in the Alaska Coasts, East Coast, Pacific Islands, and West Coast are considered, and historical data for $30$ years, collected from $1976$ through $2005$, is used to estimate the joint probability distribution of waves for each location as a function of significant wave heights $H_{s}$ and wave periods $T_{p}$. 
Since the collected data form a non-parametric representation of the probability density function (PDF), a kernel distribution characterized by a smoothing function and a bandwidth value is required to construct the PDF.
Using a Gaussian quadrature approach with $n_{gq}$ points in each dimension ($H_{s}$ and $T_{p}$), we obtained the Legendre-Gauss nodes and weights \citep{golub1969calculation}.
Then, for each year of the study, the PDF of waves was constructed using these points in \texttt{Matlab}'s \texttt{ksdensity} function. 
The sites of interest, along with their joint PDF for the first year of the study, are shown in Fig.~\ref{fig:Wave_loc}.

The sea state is described using the JOint North Sea WAve Project (JONSWAP) spectrum, defined as:
\begin{align}
    \label{eqn:JS}
    S_{JS} (H_{s},T_{p},\omega) = \alpha_{s} \omega^{-5} \exp{\left[ -\beta_{s} \omega^{-4}\right]} 
\end{align}
\noindent
where $\omega$ is the angular frequency, and $\alpha_{s}$ and $\beta_{s}$ are parameters of the spectrum defined as:
\begin{align}
    \label{eqn:JS_parameters_AlphaBeta}
    \alpha_{s} &= \dfrac{\beta_{s}}{4}H_{s}^{2}C(\gamma)\gamma^{r}\\ 
    \beta_{s} &=\dfrac{5}{4}\omega_{p}^{4}   
\end{align}
\noindent
where $\omega_{p}$ is the peak angular frequency, and $C(\cdot)$ is a normalizing factor calculated as:
\begin{align}
    \label{eqn:JS_parameters_C}
    C(\gamma) &= 1 - 0.287 \ln(\gamma) 
\end{align}
\noindent
In these equations, $\gamma$ is defined as:
\begin{align}
    \label{eqn:JS_parameters_gamma}
    \gamma = 
    \begin{cases}
         5~ &\textrm{for}~~ \dfrac{T_{p}}{\sqrt{H_{s}}} \leq 3.6\\
         \exp{(5.75 -1.15\dfrac{T_{p}}{\sqrt{H_{s}}})}  &\textrm{for}~~ 3.6 \leq \dfrac{T_{p}}{\sqrt{H_{s}}} \leq 5 \\
         1 &\textrm{for}~~ \dfrac{T_{p}}{\sqrt{H_{s}}} > 5\\
    \end{cases}
\end{align}
\noindent
Finally, $r$ is defined as:
\begin{align}
    \label{eqn:JS_parameters_r}
    r = \exp{\left[\dfrac{-1}{2\sigma^{2}}\left(\dfrac{\omega}{\omega_{p}} - 1\right)^2\right]} \quad 
    \textrm{where}~\sigma = 
    \begin{cases} 
        0.07~ \textrm{for} ~\omega \leq \omega_{p}\\
        0.09~ \textrm{for}~ \omega > \omega_{p}
    \end{cases}
\end{align}
\noindent
More details on JONSWAP spectrum can be found in \cite{ning2022modelling}.

The incident wave field can be modeled by irregular waves constructed through the superposition of $n_{r}$ regular waves.
With the assumption of $\beta_{w} = 0$ for the angle of wave direction, the irregular incident wave is approximated as: 
\begin{align}
    \label{eqn:irregularwave}
    \eta (x,t) = \sum_{i=1}^{n_{r}} \dfrac{H_{i}}{2}\cos{(k_{i}x - \omega_{i} t + \phi_{i})}
\end{align}
\noindent
where $H_{i}$ is the wave amplitude, $\phi_i$ is the randomly generated wave phase, and $k$ is the wave number satisfying the dispersion relation:
\begin{align}
    \label{eqn:dispersion}
    \omega^{2} = gk\tanh{kh} \approx 
    \begin{cases}
        gk~ &\textrm{as}~~ kh \to \infty\\
        gk^{2}h &\textrm{as}~~ kh \to 0
    \end{cases}
\end{align}
\noindent
where $g$ is the gravitational acceleration, and $h$ is the water depth.

\subsection{Wave-Structure Interactions} 
\label{subsec:Wave_Structure_Interactions}

In this section, we briefly describe the basics of wave-structure interactions.
For further details, interested readers are referred to \cite{ning2022modelling, folley2016numerical, falnes2020ocean}.

\subsubsection{Linear Potential Flow Theory}
\label{subsubsec:Linear_Potential_Flow_Theory}
Ocean waves are generally described through the Navier-Stokes equations, which, along with continuity equation and incompressibility constraint, form a system of nonlinear partial differential equations \citep{ning2022modelling}. 
While the numerical solution of this system of equations through computational fluid dynamics (CFD) is computationally expensive, making simplifying assumptions (such as incompressible, inviscid, and irrotational flow) enables the usage of potential flow theory, which is characterized by the Laplace equation, and boundary constraints at the free surface and any rigid boundary.
Under the assumption of non-steep waves, the free-surface boundary constraints can be linearized, resulting in linear potential theory, also known as Airy wave theory \citep{ning2022modelling}.

\subsubsection{Hydrodynamic and Hydrostatic Forces}
In linear potential flow theory, the fluid velocity potential $\Phi$ has contributions associated with incident $\Phi_{i}$, scattered $\Phi_{s}$, and radiated waves $\Phi_{r}$:
\begin{align}
    \Phi = \Phi_{i} + \Phi_{s} + \Phi_{r}
\end{align}
The contribution from incident waves considers the propagation of the wave in the absence of any structure.
The interference resulting from the interaction of waves with a motionless structure is characterized through scattered contributions, and radiated waves present contributions from the motion of the structure. 

The force and force moment on a structure are separated into forces resulting from the incident and scattered waves.
These quantities are referred to as the excitation and radiation force, respectively.
Since the excitation force results from the interaction between incident waves and motionless structures, it captures two effects: the motion of the incident wave, also known as the Froude-Krylov force, as well as the scattering effect. 
The radiation force results from the device's own oscillatory motion in the absence of an incident wave field, with the real and imaginary parts constituting the added mass and the damping coefficient, respectively.

While the incident wave travels only in one direction, the scattered and radiated waves propagate along all directions, thereby affecting all WEC devices, regardless of their location.
Therefore, these waves redistribute part of the incident energy in all directions \citep{babarit2013park}, making them of significant importance in the layout optimization of WECs. 
Hydrostatic forces, which can be estimated analytically or numerically, result from the change in the hydrostatic pressure on the wetted surface of the body as it moves from its equilibrium position.

\subsection{Multiple Scattering}
MS is an efficient numerical scheme that uses an iterative procedure in order to calculate the total potential field for a given array configuration.
Each iteration involves the estimation of the diffraction and radiation coefficients for each isolated body, considering the added potential from the previous iterations and subject to each body's prescribed boundary conditions.
These iterations continue up to a prescribed scattering order $m_{\text{MS}}$.
While MS is more efficient than BEM solvers \citep{folley2016numerical}, the computational cost increases linearly with the maximum scattering order and quadratically with the number of WECs in the array \citep{zhang2020surrogate}.
Therefore, direct usage of MS in the optimization loop can become computationally prohibitive for large array studies.

\subsection{Dynamics and Control of WECs} 
\label{subsec:Dynamics_of_WEC_Arrays}
When the device exhibits small amplitudes in its oscillatory motions, the hydrodynamic interactions between a WEC device and ocean waves can be simplified.
Linear representation of forces involved in WEC dynamics can be utilized to carry out the modeling in the frequency domain.
Despite their limitations, frequency-domain models are the first design space that engineers thoroughly investigate to gain early-stage insights before moving on to the more computationally-expensive time-domain models.
Due to their efficiency, frequency-domain models are particularly suitable for concurrent, system-level design investigations. 
This section briefly presents the formulations for the dynamics of WEC arrays in the frequency domain.

Considering regular waves with radial frequency of $\omega$ and unit amplitude as an input, the equations of motion for $\Nwec$ buoys in the frequency domain can be described as:
\begin{align}
\label{eqn:EquationofMotion1}
     -{\omega}^{2} \Mass \hat{\bm{\xi}}\OMF = \hat{\Force}_{\text{FK}}\OMF + \hat{\Force}_{\text{s}}\OMF + \hat{\Force}_{\text{r}}\OMF +  \hat{\Force}_{\text{hs}}\OMF + \hat{\Force}_{\text{pto}}\OMF
\end{align}
\noindent
where $\hat{\parm}$ is the complex amplitude of $\parm$, $\hat{\bm{\xi}}(\cdot) \in \mathbb{R}^{\Nwec \times 1}$ is the displacement vector, $\hat{\Force}_{\text{FK}}(\cdot)$ is the Froude-Krylov force, $\hat{\Force}_{\text{s}}(\cdot)$ is the scattering force vector, $\hat{\Force}_{\text{r}}(\cdot)$ is the radiation force, $ \hat{\Force}_{\text{hs}}(\cdot)$ is the hydrostatic force, $\hat{\Force}_{\text{pto}}(\cdot)$ is the power-take-off (PTO) force, all defined in $\in \mathbb{R}^{\Nwec \times 1}$.
Here, $\Mass \in \mathbb{R}^{\Nwec \times \Nwec}$ is the diagonal mass matrix.
The excitation force is defined as:
\begin{align}
\label{eqn:F_excitation}
    \hat{\Force}_{\text{e}}\OMF = \hat{\Force}_{\text{FK}}\OMF + \hat{\Force}_{\text{s}}\OMF
\end{align}
\noindent
The radiation force, which is calculated as a function of hydrodynamic damping coefficient matrix $\DampingCoeff(\cdot)$, and added mass coefficient matrix $\AddedMass(\cdot)$, is mathematically described as:
\begin{align}
\label{eqn:F_radiation}
    \hat{\Force}_{\text{r}}\OMF = -i{\omega} \DampingCoeff\OMF\hat{\bm{\xi}}\OMF + {\omega}^{2}\AddedMass\OMF\hat{\xi}\OMF
\end{align}
\noindent
where $\DampingCoeff(\cdot) \in \mathbb{R}^{\Nwec \times \Nwec}$ and $\AddedMass(\cdot)\in \mathbb{R}^{\Nwec \times \Nwec}$ capture the energy transmitted from WEC motions to the water, and inertial increase due to water displacement as a result of the WEC motion, respectively \citep{folley2016numerical}.
For $\Nwec$ WEC devices in a fixed array with only heave motion, $\AddedMass\OMF$ and $\DampingCoeff\OMF$ matrices are \citep{lyu2019optimization}:
\begin{align}
    \label{eqn:AandBMatrices}
    \AddedMass\OMF &= \begin{bmatrix}
        \addedmassv_{11}\OMF&~~ \addedmassv_{12}\OMF&~~  \cdots&~~ \addedmassv_{1\Nwec}\OMF&\\
        \addedmassv_{21}\OMF&~~ \addedmassv_{22}\OMF&~~  \cdots&~~ \addedmassv_{2\Nwec}\OMF&\\
        \vdots& \vdots&  \vdots&  \vdots \\
        \addedmassv_{\Nwec1}\OMF& \cdots&  \cdots& \addedmassv_{\Nwec\Nwec}\OMF&
    \end{bmatrix}\\
        \DampingCoeff\OMF &= \begin{bmatrix}
        \dampingcoeffv_{11}\OMF&~~ \dampingcoeffv_{12}\OMF&~~ \cdots&~~ \dampingcoeffv_{1\Nwec}\OMF&\\
        \dampingcoeffv_{21}\OMF&~~ \dampingcoeffv_{22}\OMF&~~ \cdots&~~ \dampingcoeffv_{2\Nwec}\OMF&\\
        \vdots& \vdots& \vdots&  \vdots \\
        \dampingcoeffv_{\Nwec1}\OMF& \cdots& \cdots& \dampingcoeffv_{\Nwec\Nwec}\OMF&
    \end{bmatrix}
\end{align}
The hydrostatic force, resulting from the balance between buoyancy and gravity, is calculated as:
\begin{align}
    \hat{\Force}_{\text{hs}}\OMF = - G \hat{\bm{\xi}}\OMF
\end{align}
\noindent
where $G = \rho gS$ is the hydrostatic coefficient, with $g$ being the gravitational constant, and $S$ being the cross-sectional area at the undisturbed sea level calculated as a function of the WEC radius: $ \pi \Rwec^2$ for a heaving cylinder. 

The PTO system, which converts mechanical motion into electricity, has significant implications on the sizing and economic performance of WECs \citep{tan2021influence, tan2020feasibility}.
Incorporating an active control strategy significantly increases the power captured by WECs \citep{herber2013wave, tedeschi2010analysis, clement2012discrete}.
Reactive control, which is one of the earliest control strategies developed for WEC devices, uses a bidirectional power flow between the PTO spring and the buoy \citep{ning2022modelling} and is described in linear form:
\begin{align}
    \hat{\Force}_{\text{pto}}({\omega}) = -i{\omega}\BPTO\hat{\bm{\xi}}{(\omega)} - \KPTO \hat{\bm{\xi}}{(\omega)}
\end{align}
\noindent
where $\KPTO \in \mathbb{R}^{\Nwec \times \Nwec}$ and $\BPTO \in \mathbb{R}^{\Nwec \times \Nwec}$ are diagonal matrices of stiffness and damping of PTO systems for WEC devices, respectively.
The complex amplitude of the motions of all WECs for a regular wave of frequency $\omega$ and unit amplitude can now be described as a transfer function matrix:
\begin{align}
{\hat{\bm{\xi}}(\omega)} &= {\mathbf{H}(\omega)\hat{\Force}_{\text{e}}({\omega})} \label{eq:TF} \\
\mathbf{H}(\omega) &= \left [ [{\omega}^2(\Mass+\AddedMass({\omega})) + G + \KPTO] + i{\omega}(\DampingCoeff({\omega}) + \BPTO)  \right ]^{-1} \label{eq:TF_denom}
\end{align}

The time-averaged absorbed mechanical power for a sea state with significant wave height of $H_{s}$ and peak period of $T_{p}$ can then be described as:
\begin{align}
    {\mathbf{p}_{m}(H_{s}, T_{p}, \omega) = \frac{1}{2}{\omega}^2 \hat{\bm{\xi}}^{T}\BPTO \hat{\bm{\xi}} }  \label{eq:pto_power1}
\end{align}
\noindent
For year $y$, the absorbed power in each desired sea state is estimated by integrating the product of the wave spectrum with the time-averaged power of Eq.~(\ref{eq:pto_power1}) over all frequencies \citep{neshat2022layout, borgarino2012impact}:
\begin{align}
    {\mathbf{p}_{i}(H_{s}, T_{p}, y) = \int_{0}^{\infty}S_{JS}(H_{s},T_{p},\omega)\mathbf{p}_{m}(H_{s}, T_{p}, \omega)\textrm{d}\omega } \label{eq:pto_power2}
\end{align}
\noindent
where $\mathbf{p}_{i}(H_{s}, T_{p}, y)$ is the mechanical power matrix.
Considering all sea states, this equation can be estimated using \citep{mercade2017layout}:
\begin{align}
    {\mathbf{p}_{i}(H_{s}, T_{p}, y)} &= {\sum_{k = 0}^{n_{w}}  \Delta\omega_{k} S_{JS}(H_{s},T_{p},\omega_{k})\mathbf{p}_{m}(H_{s}, T_{p}, \omega_{k})}  \label{eq:pto_power3} 
\end{align}
\noindent
where $n_{w}$ is the number of frequencies in the discretized form.
To estimate power production over the entire farm lifetime ($y=1,2, \cdots ,n_{yr}$), the associated annual probability matrices are utilized \citep{neary2014methodology}:
\begin{align}
    {p_{a} = \eta_{\text{pcc}}\eta_{\text{oa}}\eta_{\text{t}}\sum_{y = 1}^{n_{yr}}\mathbf{p}_{i}(H_{s}, T_{p}, y) \mathbf{p}_{r}(H_{s},T_{p}, y)}
\end{align}
\noindent
where $p_{a}$ is the average power, $\mathbf{p}_{r}(H_{s},T_{p}, y)$ is the joint probability distribution of the wave climate in the $y$th year, $\eta_{\text{pcc}}$ is the efficiency of the power conversion chain, $\eta_{\text{oa}}$ is the operational availability, and $\eta_{\text{t}}$ is the transmission efficiency. 
Average power per unit volume of the WEC device, described by $p_{v}$, is used as the objective function \citep{falnes2020ocean}:
\begin{align}
    \label{eq:powerpervolumeObj}
    p_{v} = \frac{p_{a}}{\pi \Rwec^2 \Dwec} 
\end{align}
\noindent
where $\Rwec$ and $\Dwec$ are the radius and draft of the heaving cylinder WEC device, respectively.

\subsection{Array Considerations} 
\label{subsec:ArrayConsiderations}

\begin{figure}
    \centering
    \includegraphics[width = \columnwidth]{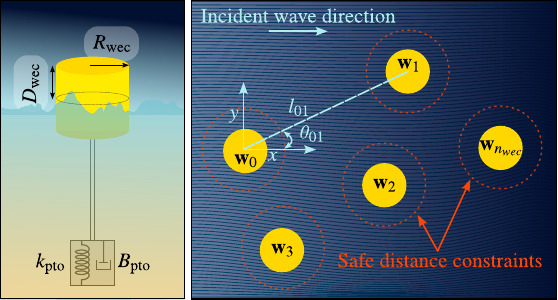}
    \caption{Illustration of WEC and its layout considerations.}
    \label{fig:WaveFarm}
\end{figure}

A WEC farm, consisting of $\Nwec$ WEC devices is considered and characterized by the $2-$by-$\Nwec$ dimensional layout matrix $\AL = [\mathbf{w}_{1}, \mathbf{w}_{2}, \cdots, \mathbf{w}_{\Nwec}]$.
Each element of $\Nwec$ is composed of a vector of the center coordinates of each body in the Cartesian coordinate system.
As an example, the $p$th WEC is characterized by $\mathbf{w}_{p} = [x_{p}, y_{p}]^{T}$.
The relative distance between two WECs, namely $p$th and $q$th devices, and their relative angle with respect to the positive direction of the $x$-axis are described as $\Dispq$ and $\Angpq$, respectively.
The incident wave angle $\beta_{w}$ is measured relative to the positive direction of $x$-axis.
Since the coordinate system may be rotated for these axisymmetric bodies, the incident wave angle may be set to $0$ without any loss of generality \citep{zhang2020surrogate}.
Figure \ref{fig:WaveFarm} illustrates WECs in an array.

\begin{figure}[]
    \centering
     \includegraphics[width=\columnwidth]{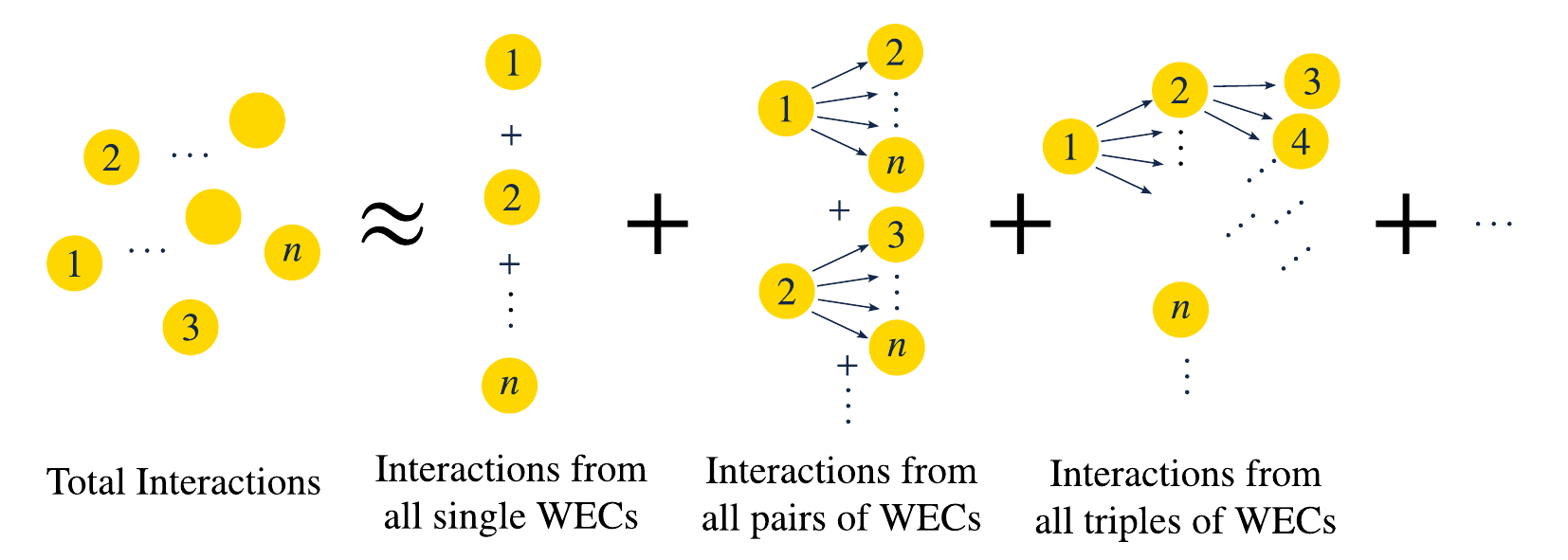}
    \caption{Illustration of many-body expansion principle.}
    \label{fig:MBE}
\end{figure}

\subsection{Surrogate Models for Hydrodynamic Interactions}
\label{subsec:SM_HIE}

With the goal of improved computational efficiency, surrogate models provide a simplified approximation of input/output relationships.  
Surrogate models in general, and artificial neural networks (ANNs) in particular have been used in the design of WEC devices \citep{anderlini2017reactive,thomas2018experimental,valerio2008identification,zhang2020surrogate,li2023physics}.
Among these, ~\cite{zhang2020surrogate} developed surrogate models to capture the hydrodynamic interactions effect within a WEC farm using both Gaussian process regression and many-body expansion principles.
In this work, we follow the path of \cite{zhang2020surrogate} to develop surrogate models that can estimate the hydrodynamic interaction effect using artificial neural networks and the principles of many-body expansion (MBE).

\subsubsection{Artificial Neural Networks.}
\label{subsec:ANN}
Biologically inspired by the human brain, ANNs are an interconnected network of nodes or neurons in a layered structure. 
Every node has its associated weight, bias, and transfer function (such as the logistic function) that allows each neuron to output a value between $0$ and $1$ based on the weighted sum of its inputs.
The weights are found by learning on the training data through an algorithm that computes the gradient vector to minimize a prediction error \citep{lecun2015deep} through a performance measure, such as the mean-squared error. 
Network structure and parameters may be adjusted to ensure that ANN performs reasonably well in predicting the outputs. 
For more information about ANN, the readers are referred to \cite{haykin1998neural}.
In this article, we utilize feedforward shallow neural networks.
While calibrating the ANN may take some effort, the resulting surrogate models can predict the desirable outputs with high accuracy and efficiency and, thus, facilitate the effective implementation of layout optimization for WECs.

\subsubsection{Many-body Expansion.}
\label{subsec:MBE}
In many-body expansion (MBE), the total interaction effect among $\Nwec$ bodies is estimated as a summation of effects corresponding to a finite number of clusters. 
These clusters are systematically selected to capture the effects of a single, two-, three-, and $\NMBE$-bodies \citep{suarez2009thermochemical}. 
Representing the desired hydrodynamic interaction effect with $\psi(\AL)$, MBE up to $\NMBE$ clusters can be estimated as \citep{zhang2020surrogate}:
\begin{align}
    \label{eqn:MBE1}
    \psi(\AL) &\approx \sum_{i=1}^{\Nwec} \psi(\MyBold{w}_{i}) + \sum_{i=1}^{\Nwec-1} \sum_{j>i}^{\Nwec}\Delta\psi(\MyBold{w}_{i},\bm{w}_{j})  + \cdots \\
    & \quad + \sum_{i=1}^{\Nwec-2}\sum_{j>i}^{\Nwec-1}\sum_{k>j}^{\Nwec} \Delta\psi(\MyBold{w}_{i},\MyBold{w}_{j}, \MyBold{w}_{k}) + \cdots \notag \\
    & \quad + \sum_{i=1}^{\Nwec-\NMBE}\cdots\sum_{k>j}^{\Nwec} \Delta\psi(\MyBold{w}_{i}, \cdots, \MyBold{w}_{k}) \notag
\end{align}
\noindent
Using an alternative notation $\{ \AL\}_{l}^{\NMBE}$ to represent the $l$th distinct $m$-body cluster, this formulation can be written as:
\begin{align}
\label{eqn:MBE2}
    \psi(\AL) &\approx \sum_{l=1}^{\Nwec} \psi(\{ \AL_{l}^{1}\}) + \sum_{l=1}^{\Nwec!/[2!(\Nwec-2)!]} \Delta\psi(\{ \AL\}_{l}^{2}\})  + \cdots \\
    & \quad + \sum_{l=1}^{\Nwec!/[3!(\Nwec-3)!]} \Delta\psi(\{ \AL\}_{l}^{3}\}) + \cdots \notag\\
    & \quad + \sum_{l=1}^{\Nwec!/[\NMBE!(\Nwec-\NMBE)!]} \Delta\psi(\{ \AL\}_{l}^{\NMBE}\}) \notag
\end{align}
\noindent
where $\Delta\psi(\cdot)$ is the additive interaction effect, calculated as:
\begin{align}
\label{eqn:MBEadditive}
    \Delta\psi(\{\AL\}_{l}^{\NMBE}) &= \psi(\{\AL\}_{l}^{\NMBE}) - \left [ \sum_{r=1}^{\NMBE} \psi(\{ \{ \AL\}_{l}^{\NMBE} \}_{r}^{1})  \right.  + \cdots \\
    &\left. + \sum_{r=1}^{\NMBE!/[2!(\NMBE-2)!]} \Delta\psi(\{ \{ \AL\}_{l}^{\NMBE} \}_{r}^{2}) \right. + \cdots \notag\\
    & \left.+ \sum_{r=1}^{\NMBE} \Delta\psi(\{ \{ \AL\}_{l}^{\NMBE} \}_{r}^{\NMBE-1}) \right] \notag
\end{align}
\noindent
In this study, we only consider interaction effects of two-body clusters. 
Therefore, for each pair of $p$ and $q$ WECs, the additive effect can be written as:
\begin{align}
\label{eqn:MBEadditive2}
    \Delta\psi(\MyBold{w}_{p}, \MyBold{w}_{q}) &= \psi(\MyBold{w}_{p}, \MyBold{w}_{q}) - \psi(\MyBold{w}_{p})-\psi(\MyBold{w}_{q})
\end{align}
\noindent
Figure~\ref{fig:MBE} shows the general concept of MBE used up to second order in this study.

\section{Constructing Surrogate Models}
\label{sec:Constructing_Surrogate_Models}

In this section, we discuss some considerations in constructing the surrogate models, including input/output structure, sampling strategies, and preparing the training data.
We also demonstrate the capability of the resulting models to estimate the objective function with reasonable accuracy.

\subsection{Surrogate Model Structure}
\label{subsec:Developing_Surrogate_Models}

In order to generate the training data for ANNs, the first step is to identify inputs and outputs.
In this article, we are interested in estimating radiation and excitation forces exerted on the WEC for all one- and two-WEC clusters. 
This results in frequency-dependent output structures with components associated with the added mass, damping coefficient, and the real and imaginary parts of the excitation force.
These quantities of interest (QoI) constitute the output vector of our ANNs and are normalized according to the following relationships:
\begin{align}
\tilde{\Force}_{\text{e}}(\omega) &= {\Force}_{\text{e}}(\omega) /(\rho g \pi\Rwec^2 \Dwec) \\
\tilde{\AddedMass}(\omega) &= \AddedMass(\omega)/(\rho \pi \Rwec^3)\\
\tilde{\DampingCoeff}(\omega) &= \DampingCoeff(\omega) / (\omega \rho \pi \Rwec^3)
\end{align}
\noindent
While slightly different from the conventional normalization factors found in the literature \citep{zhang2020surrogate, mavrakos1987hydrodynamic, herber2013wave}, these equations are more effective for the training of ANNs because they better scale the QoIs by considering input dependencies.
In this article, we develop an individual model for each of the desired outputs in order to simplify the task of tuning network parameters.

For the single-body cluster, the input vector includes the radius $\Rwec$, and slenderness ratio (i.e.,~radius over draft) $\RDwec$ of the WEC.
The hydrodynamic interaction effect then needs to be calculated for samples from the input space only at the center of the coordinate system, i.e.,~$\MyBold{w}_{0} =[0,0]^{T}$. 
This outcome is because the radiation output from these surrogate models is independent of the location of the WEC (i.e.,~$\tilde{\AddedMass}(\AL) = \tilde{\AddedMass}([0, 0]^{T})$ and $\tilde{\DampingCoeff}(\AL) = \tilde{\DampingCoeff}([0, 0]^{T})$).
However, it depends on plant specifications, such as WEC radius and slenderness ratio.
The excitation force output is invariant with respect to translations along the $y$-axis and is symmetric with respect to the wave propagation direction; however, for translation by $L$ along the $x$ direction, a phase shift is created by $\exp{(-ikL)}$, where $k$ is the wave number (from Eq.~(\ref{eqn:dispersion})), and $i=\sqrt{-1}$.

The input vector for the one-body cluster is defined as $\bm{v}_{1} = [\Rwec, \RDwec]^{T}$.
The outputs are comprised of elements associated with the normalized added mass $\tilde{\addedmassv}(\omega)$, damping coefficient $\tilde{\dampingcoeffv}(\omega)$, and real and imaginary parts of the excitation force $\tilde{f}_{\text{e}}(\omega)$.
These quantities are represented in the output vector $\tilde{\bm{y}}_{1} = [\tilde{\addedmassv}, \tilde{\dampingcoeffv}, \text{Re}\{\tilde{f}_\text{e}\}, \text{Im}\{\tilde{f}_\text{e}\}]^{T}$.
The input-output relationship, captured through the resulting surrogate models, is then represented as: 
\begin{align}
\label{eq:SM1}
    \tilde{\bm{y}}_{1} = \bm{f}_{1}({\bm{v}}_{1}).
\end{align}
\noindent
where $\bm{f}_{1}$ is the vector of resulting ANN functions for the single-body cluster, composed of $\bm{f}_{1} = [f^{a}_{1}, f^{b}_{1}, f^{f_{r}}_{1}, f^{f_{im}}_{1} ]^{T}$.

The interaction effect among two-body clusters is dependent on an input vector that, in addition to WEC radius and slenderness ratio, includes the relative distance $\Dispq$ and angle $\Angpq$ between the two bodies.
This input vector is described as $\bm{v}_{2} = [\Rwec, \RDwec, \Dispq, \Angpq]^{T}$. 
The two-body system radiation effect depends strictly on the relative distance of the bodies, while the excitation force also depends on the relative angle.
Due to symmetry with respect to the $x$-axis, the excitation force needs to be investigated only in the range of $[0,~\pi]$.
Any other angle can be mapped to that solution \citep{zhang2020surrogate}.
Instead of directly developing the models for the additive effect described in Eqs.~(\ref{eqn:MBEadditive})--(\ref{eqn:MBEadditive2}), it is more straightforward to structure the two-body surrogate models with direct outputs from multiple scattering.  
Any required transformation, such as those described in Eqs.~(\ref{eqn:MBEadditive})--(\ref{eqn:MBEadditive2}), can be performed once the models are developed. 
These outputs are defined as 
$\tilde{\bm{y}}_{2} = [\tilde{\addedmassv}_{11}, \tilde{\addedmassv}_{12},\tilde{\dampingcoeffv}_{11}, \tilde{\dampingcoeffv}_{12},\text{Re}\{\tilde{f}_{\text{e}_{1}}\}, \text{Im}\{\tilde{f}_{\text{e}_{1}}\}]^{T}$,
where the indices $\parm_{11}$ and $\parm_{12}$ correspond to the interaction effect between any pair of WECs.
The two-body cluster surrogate models can be defined as:
\begin{align}
    \tilde{\bm{y}}_{2} = \bm{f}_{2}({\bm{v}}_{2})
\end{align}
\noindent
where $\bm{f}_{2}$ is the vector of resulting ANN functions for the two-body cluster, composed of $\bm{f}_{2} = [f^{a_{11}}_{2}, f^{a_{12}}_{2}, f^{b_{11}}_{2},f^{b_{12}}_{2}, f^{f_{r}}_{2}, f^{f_{im}}_{2} ]^{T}$.
From here, the additive effect from the two-body cluster associated with the radiation interaction can be defined as:
\begin{align}
    \Delta \tilde{\addedmassv}_{11} & = \tilde{\addedmassv}_{11} - \tilde{\addedmassv} =  f^{a_{11}}_{2}({\bm{v}}_{2}) - f^{a}_{1}({\bm{v}}_{1}) \label{eq:delta_a_11}\\
    \Delta \tilde{\addedmassv}_{12} & = f^{a_{12}}_{2}({\bm{v}}_{2})\label{eq:delta_a_12}\\
    \Delta \tilde{\dampingcoeffv}_{11} &= \tilde{\dampingcoeffv}_{11} - \tilde{\dampingcoeffv} = f^{b_{11}}_{2}({\bm{v}}_{2}) - f^{b}_{1}({\bm{v}}_{1}) \label{eq:delta_b_11}\\
    \Delta \tilde{\dampingcoeffv}_{12} & = f^{b_{12}}_{2}({\bm{v}}_{2}) \label{eq:delta_b_12}
    \end{align}
\noindent
For excitation force, the additive effect is captured as:
\begin{align}
    {\Delta \tilde{f}_{\text{e}1}} &= { (\tilde{f}_{\text{e}_{1}} - \tilde{f}_{\text{e}})\exp{(ikL)}} \label{eq:delta_f_11} \\
    &{= \left ([f^{f_{r}}_{2}({\bm{v}}_{2}) + if^{f_{im}}_{2}({\bm{v}}_{2})] - [f^{f_{r}}_{1}({\bm{v}}_{1}) + if^{f_{im}}_{1}({\bm{v}}_{1})] \right)\exp{(ikL)} }\notag
\end{align}
\noindent
Other quantities of interest, such as those corresponding to the additive effect of the first body on the second one, may be simply calculated by swapping the order of the bodies.

\subsection{Developing Training Data and ANNs}
\label{subsec:Data_Processing}

In addition to the normalization scheme discussed in Sec.~\ref{subsec:Developing_Surrogate_Models}, additional design-informed considerations are utilized to generate appropriate training data for developing ANNs.
While preventing us from producing impractical solutions, such considerations improve the training performance by limiting the range of outputs in the training set (which results in avoiding outliers in the output).
As an example, it was observed that extreme and unreasonable design combinations (e.g., designs with very small draft and very large radius and vice versa) have little practical value but often contribute significantly to increasing the range of the QoI.
Therefore, in preparing the training data, such design specifications are avoided. 

Experiments with multiple passive selection strategies, including Latin hypercube sampling (LHS) and gridded data (with both linearly- and logarithmically-spaced samples) revealed that an extremely high number of training samples is required for accurate prediction of hydrodynamic coefficients, resulting in high computational cost and inefficiencies in training ANNs.
This issue is because random samples contain progressively less information as the learning proceeds \citep{raychaudhuri1995minimisation}.
Using a selective sampling approach such as Query by Committee (QBC) \citep{seung1992query}, however, is expected to improve the selection strategy significantly.

QBC is an active learning strategy in which a committee of learners (models) are iteratively trained on a current data set.
The next query of points, characterized by a user-defined budget, is selected to be measured based on the maximum level of disagreement over the committee.
The diversity of models (which is the source of disagreement) may result from the random initialization of weights and bias terms of each network, or the training of each model on different sub-samples of the current data set \citep{raychaudhuri1995minimisation}.
The latter is similar to the query by bagging approach used in classification problems \citep{burbidge2007active}.
Samples with the highest level of disagreement are then selected to be evaluated and added to the current data set.
These new sample points have the highest contribution when resolving the disagreements among models.
This approach practically minimizes the data collection effort.   
For real-valued functions (e.g., regressor models), this disagreement is assessed through the learners' variance \citep{krogh1994neural}.
The new samples can be selected from a pool of candidates.
This approach is known as the pool-based selective sampling \citep{burbidge2007active}.
Alternatively, new samples can be obtained from an optimization problem in which the objective is to maximize the variance. 
In addition, one may select a single point or a batch of points to be evaluated and added to the current data set. 

\begin{vAlgorithm}[t!]{\columnwidth}{0.2cm}
\caption{Pool-based batch mode QBC}\label{alg:QBC}
\KwData{initial data set $\mathcal{D}_{0}$}
\For{each quantity of interest (QoI)}{
initialization\;
$\mathbb{E}[\text{var}] \leftarrow \infty$ \;
$\max[\text{mse}] \leftarrow \infty$ \;
$k \leftarrow 1$\;
\While{$ \epsilon_{\text{var}} \leq \mathbb{E}[\text{var}]$ $\vert \vert$ $\epsilon_{\text{mse}} \leq \max[\text{mse}] $}
{
create random, equal-sized sub-samples $\mathcal{D}_{i}$\;
randomly initialize committee of ANNs\;
train ANNs on sub-samples\;
evaluate models at all candidate points\;
select $20\%$ of points with maximum variance\;
divide points into $n_{b}$ clusters to find $x_{\text{new}}$\;
evaluate $x_{\text{new}}$ using MS and add to $\mathcal{D}_{0}$\;
compute $\mathbb{E}[\text{var}]$ and $\max[\text{mse}]$\;
    \If{$k > k_{\text{max}}$}
    {
        \text{break}\;
    }
    $k \leftarrow k+1$\;
}
$\mathbb{E}[\text{variance}] \leftarrow \infty$ \;
$\max[\text{mse}] \leftarrow \infty$ \;
}
\end{vAlgorithm}

The algorithm used for the QBC is shown in Algorithm~\ref{alg:QBC}.
In this article, we utilize a batch-mode, pool-based QBC algorithm in order to ensure an efficient selection strategy for each model.
The initial measured data set $\mathcal{D}_{0}$ was created using a combination of points inside and on the boundary of the design space.
A committee of $10$ and $5$ neural networks were randomly initialized and trained on equally-sized, randomly selected sub-samples $\mathcal{D}_{i}$ from the current set, using the mean squared error (MSE) metric, for the single WEC and two-WEC clusters, respectively.
Each individual data set was randomly divided such that $70\%$ of the data was used for training, $15\%$ for validation, and $15\%$ for testing.
Due to its observed superior performance, \texttt{Matlab}'s scaled conjugate gradient backpropagation algorithm, \textit{trainscg}, was used for the training of all networks (lines 1--9 of Algorithm~\ref{alg:QBC}).

The selection budget, also known as the batch size, was set to $n_{b} = 50$ and $n_{b}=200$ for single- and two-WEC models, respectively.
The selection of new samples $x_{\text{new}}$ from the large pool of candidates with $5\times 10^4$ and  $5\times 10^5$ points for single- and two-WEC models, respectively, entails identifying points that exhibit the largest expected variance (lines 10--11 of Algorithm~\ref{alg:QBC}).
To do this, $20\%$ of the samples with the highest variance were identified and grouped into $n_{b}$ clusters using k-means clustering \citep{likas2003global, pham2005selection} (line 12 of Algorithm~\ref{alg:QBC}).
The cluster centroid locations were then selected as new samples, evaluated, and then added to the measured data set (line 13 of Algorithm~\ref{alg:QBC}).
These steps are repeated until the stopping criteria are met (lines 14--19 of Algorithm~\ref{alg:QBC}).
Three stopping criteria were selected to create a balance between the performance of the models and the computational budget.
In addition to requiring the expected variance to drop below a threshold $\epsilon_{\text{var}}$, the maximum mean-squared error from the measured points was selected as a stopping criterion.
Finally, for developing two-WEC models, a maximum of $20$ iterations was prescribed. 

\begin{table}[t]
    \caption{Parameters of the hydrodynamic problem, QBC algorithm, and the multiple scattering approach.}
    \label{Tab:TR_settings}
    \centering
    \begin{tabular}{l c c c c}
    \hline  \hline
    \textrm{\textbf{Symbol}} & \textrm{\textbf{Definition}} & \textrm{\textbf{Value}} & \textrm{\textbf{Unit}} \\
    \hline
    $\omega$ & \textrm{frequency} & $[0.3~~ 2]$ & [\unit{rad/s}] \\
    $h$ & \textrm{water depth}&$50$ & [\unit{m}] \\
    $\beta_{w}$ & \textrm{wave angle}& $0$ & [\unit{rad}] \\
    $\Rwec$ & WEC radius & $[0.5,~10]$ & [\unit{m}] \\
    $\RDwec$ & slenderness ratio & $[0.2,~10]$ & - \\
    $\Dwec$ & WEC draft & $[0.5,~20]$ & [\unit{m}] \\
    $\Dispq$ &  relative distance & $[2\Rwec,~1000]$ & [\unit{m}]\\
    $\Angpq$ &  relative angle & $[0,~\pi]$ & [\unit{rad}] \\
    \hline
    $\text{tol}_{var}$ & variance tolerance & $10^{-3}$ & -\\
    $\text{tol}_{mse}$ & mse tolerance & $10^{-3}$& -\\
    $n_{b}$ & batch size & $50$~\&~$200$& -\\
    $n_{cmt}$ & $\#$ committee & $10$ \& $5$& -\\ 
    \hline
    $\text{MS}_m$ & main fluid order& $5$ & - \\
    $\text{MS}_n$ & fluid below body& $40$ & - \\
    $\text{MS}_j$ & diffraction  order& $15$ & - \\
    $\text{MS}_s$ & interaction order & $5$ & - \\
    \hline \hline
    \end{tabular}
\end{table}

The selected samples were evaluated through the MS approach at the constant water depth of $h = 50~ [\unit{m}]$, for the range of radial frequencies from $\omega_{0} = 0.3~[\unit{rad/s}]$ to $\omega_{f} = 2~[\unit{rad/s}]$, with a total of $100$ evenly-spaced values.
Parameters of the hydrodynamic problem, along with those associated with the algorithm setup and MS, are shown in Table~\ref{Tab:TR_settings}.
Specifically, MS parameters include the order of series truncation for the main fluid $\text{MS}_{m} = 5$, order of series truncation for the fluid below the body $\text{MS}_{n}= 40$, order of series truncation for diffraction problem $\text{MS}_{j}= 15$, and order of interaction $\text{MS}_{s}=5$.

\subsection{Validation of Surrogate Models}
\label{subsec:Validation_Surrogate_Models}

Before utilizing SMs within the optimization problem, it is necessary to assess their performance.
In this article, SMs are validated with different goals and through different means.
Overall, four validation schemes are included in this article:
\begin{enumerate}[topsep=0pt,itemsep=-1ex,partopsep=1ex,parsep=1ex,label=$\bullet$]
    \item SM output validation in Sec.~\ref{subsubsec:SMOV}
    \item Validation of objective function estimation through SMs in Sec.~\ref{subsubsec:SMOFV}
    \item Validation of optimal plant estimation through SMs in Sec.~\ref{subsubsec:CaseStudy1}
    \item  Validation of optimal layout estimation through SMs in Sec.~\ref{subsubsec:CaseStudy2}
\end{enumerate}
The first two validation schemes are discussed in this section, while the latter two which leverage an optimization solution, are discussed in Sec.~\ref{sec:Optimization}.

\subsubsection{Surrogate Model Output Validation} 
\label{subsubsec:SMOV}
The resulting ANNs exhibit an acceptable correlation coefficient (over different frequency-dependent outputs) for the training, testing, and validation sets, indicating a strong linear association. 
In order to ensure robustness in predicting outputs, all of the committee members are utilized, and their average output is consistently used as the prediction value. 
The performance of models in various regions of the design space is assessed through the MSE metric, averaged over all of the frequency-dependent outputs.
For two-WEC models, a conservative worst-case (maximum error) approach is employed to evaluate performance. 

\begin{figure*}[t]
\captionsetup[subfigure]{justification=centering}
\centering
\begin{subfigure}{0.23\textwidth}
\centering
\includegraphics[width = \columnwidth]{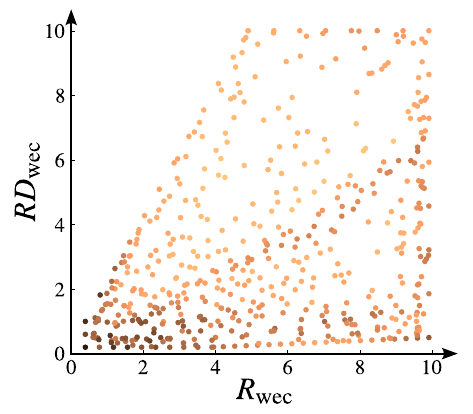}
\caption{$\tilde{\addedmassv}$.}
\label{fig:MSE_A_1WEC}
\end{subfigure}%
\begin{subfigure}{0.23\textwidth}
\centering
\includegraphics[width = \columnwidth]{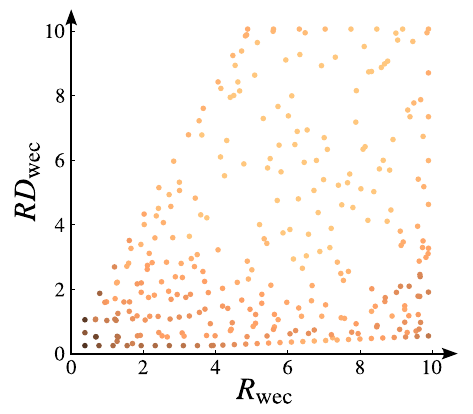}
\caption{$\tilde{\dampingcoeffv}$.}
\label{fig:MSE_B_1WEC}
\end{subfigure}%
\begin{subfigure}{0.23\textwidth}
\centering
\includegraphics[width = \columnwidth]{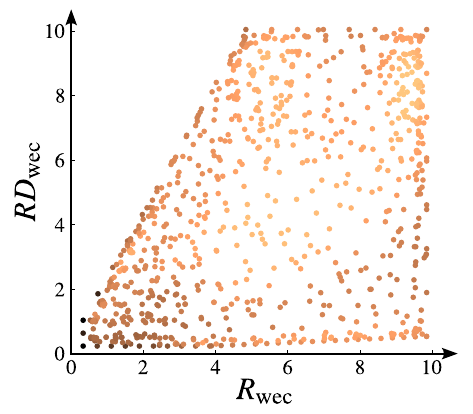}
\caption{$\text{Re}\{\tilde{f}_\text{e}\}$.}
\label{fig:MSE_F_r_1WEC}
\end{subfigure}%
\begin{subfigure}{0.23\textwidth}
\centering
\includegraphics[width = \columnwidth]{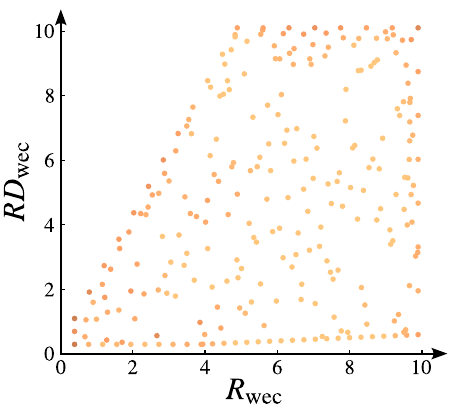}
\caption{$\text{Im}\{\tilde{f}_\text{e}\}$.}
\label{fig:MSE_F_i_1WEC}
\end{subfigure}%
\begin{subfigure}{0.08\textwidth}
\centering
\includegraphics[scale = 0.5]{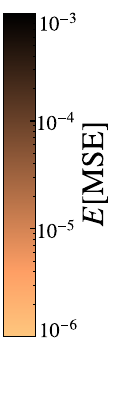}
\caption*{}
\label{fig:MSE_1WEC_LEG}
\end{subfigure}
\captionsetup[figure]{justification=centering}
\centering
\caption{ANN model performance for estimation of various hydrodynamic coefficients of a single WEC, characterized in the plant design space through expected MSE. A different number of samples is intelligently selected for each model.} 
\label{fig:MSE_WEC}
\end{figure*}

\begin{figure*}[t]
\captionsetup[subfigure]{justification=centering}
\centering
\begin{subfigure}{0.6\textwidth}
\centering
\includegraphics[width = \columnwidth]{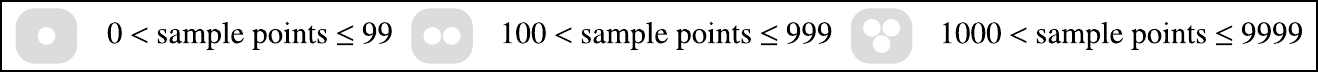}
\label{fig:Two_WEC_leg}
\end{subfigure}

\vspace{-0.75\baselineskip}

\begin{subfigure}{0.23\textwidth}
\centering
\includegraphics[width = \columnwidth]{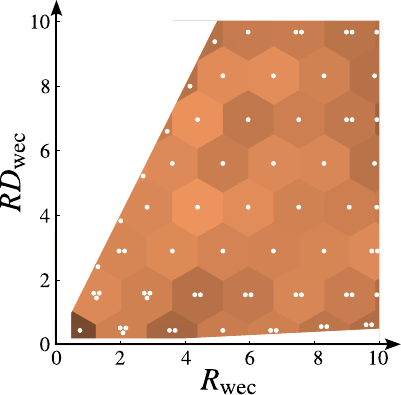}
\caption{$\tilde{\addedmassv}_{11}$.}
\label{fig:MSE_A_2WEC_Mean}
\end{subfigure}%
\begin{subfigure}{0.23\textwidth}
\centering
\includegraphics[width = \columnwidth]{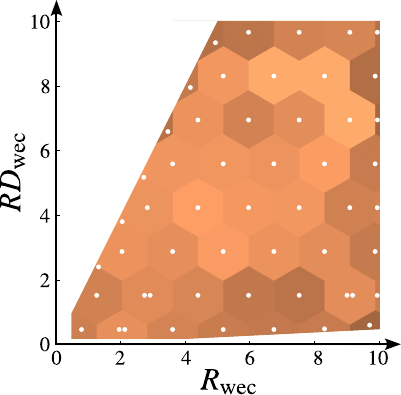}
\caption{$\tilde{\dampingcoeffv}_{11}$.}
\label{fig:MSE_B_2WEC_Mean}
\end{subfigure}%
\begin{subfigure}{0.23\textwidth}
\centering
\includegraphics[width = \columnwidth]{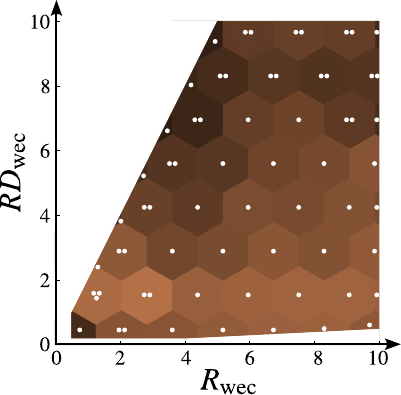}
\caption{$\text{Re}\{\tilde{f}_{\text{e}_{1}}\}$.}
\label{fig:MSE_F_r_2WEC_Mean}
\end{subfigure}%
\begin{subfigure}{0.23\textwidth}
\centering
\includegraphics[width = \columnwidth]{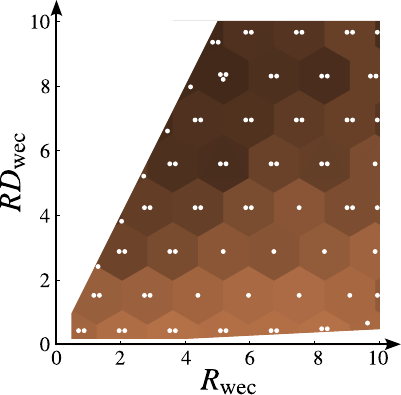}
\caption{$\text{Im}\{\tilde{f}_{\text{e}_{1}}\}$.}
\label{fig:MSE_F_i_2WEC_Mean}
\end{subfigure}%
\begin{subfigure}{0.08\textwidth}
\centering
\includegraphics[scale = 0.58]{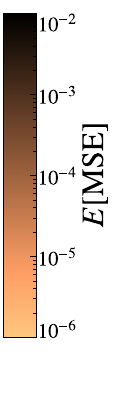}
\label{fig:MSE_2WEC_LEG_Mean}
\caption*{}
\end{subfigure}
\begin{subfigure}{0.23\textwidth}
\centering
\includegraphics[width = \columnwidth]{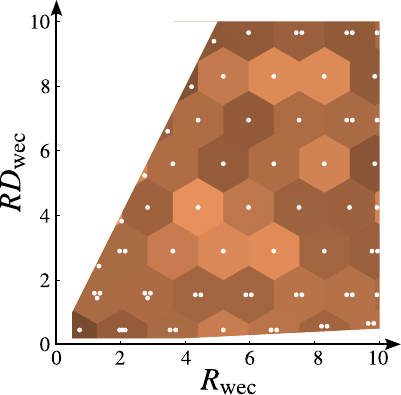}
\caption{$\tilde{\addedmassv}_{11}$.}
\label{fig:MSE_A_2WEC_WC}
\end{subfigure}%
\begin{subfigure}{0.23\textwidth}
\centering
\includegraphics[width = \columnwidth]{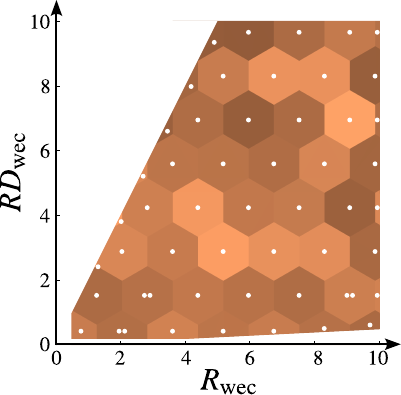}
\caption{$\tilde{\dampingcoeffv}_{11}$.}
\label{fig:MSE_B_2WEC_WC}
\end{subfigure}%
\begin{subfigure}{0.23\textwidth}
\centering
\includegraphics[width = \columnwidth]{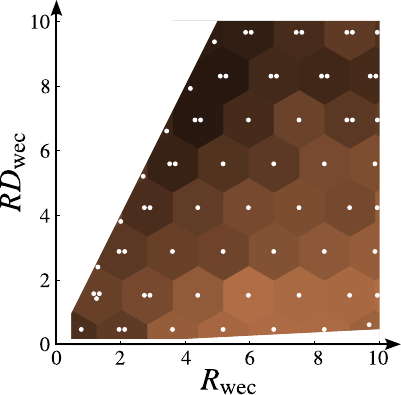}
\caption{$\text{Re}\{\tilde{f}_{\text{e}_{1}}\}$.}
\label{fig:MSE_Fe_r_2WEC_WC}
\end{subfigure}%
\begin{subfigure}{0.23\textwidth}
\centering
\includegraphics[width = \columnwidth]{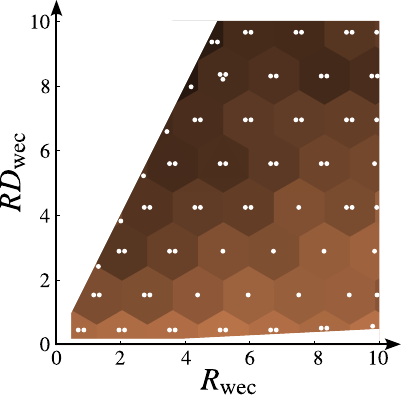}
\caption{$\text{Im}\{\tilde{f}_{\text{e}_{1}}\}$.}
\label{fig:MSE_Fe_i_2WEC_WC}
\end{subfigure}%
\begin{subfigure}{0.08\textwidth}
\centering
\includegraphics[scale = 0.58]{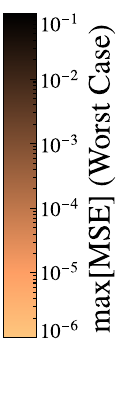}
\caption*{}
\label{fig:MSE_2WEC_LEG_WC}
\end{subfigure}
\captionsetup[figure]{justification=centering}
\centering
\caption{ANN model performance for estimation of various hydrodynamic coefficients for a two-WEC farm, characterized in the plant design space through both expected MSE (top row) and worst-case maximum error (bottom row).
A different number of samples is intelligently selected for each model. 
The four-dimensional input data (including relative distance and angle) is reduced to projections on a two-dimensional plane.
All neighboring data points are placed inside a hexagon using \cite{hex}.}
\label{fig:MSE_2WEC}
\end{figure*}

According to Fig.~\ref{fig:MSE_WEC}, the single-WEC ANN models have an acceptably small error of less than $10^{-3}$, over the entire outputs across the entire (plant) design space.
The difference in the number of samples in each case highlights some of the benefits of utilizing the QBC approach in its intelligent sampling strategy.
Due to the high dimensionality of the input data for two-WEC models, projections on the two-dimensional plant design space are used to create effective visualizations.
As shown in Fig.~\ref{fig:MSE_2WEC}, in this approach, all neighboring points are placed inside a hexagon, and the mean MSE and worst-case error (max MSE) are used to assess performance.
In Fig.~\ref{fig:MSE_2WEC}, the number of circles inside each hexagon points to the order of the number of samples in that hexagon.
It is clear that certain locations within the design space require a higher number of samples. 
In addition, according to these figures, a maximum average MSE of $10^{-2}$ and a maximum worst-case error of $10^{-1}$ can be established for these surrogate models.

\subsubsection{Objective Function Validation}
\label{subsubsec:SMOFV}
\begin{figure}[ht]
\captionsetup[subfigure]{justification=centering}
\centering
\begin{subfigure}{0.5\columnwidth}
\centering
\includegraphics[scale=\twofigscale]{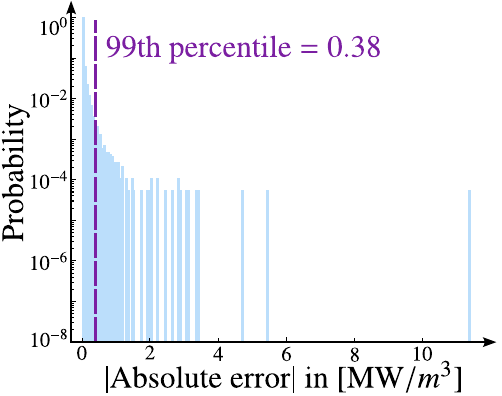}
\caption{Absolute error histogram.}
\label{fig:error_abs_obj}
\end{subfigure}%
\begin{subfigure}{0.5\columnwidth}
\centering
\includegraphics[scale=\twofigscale]{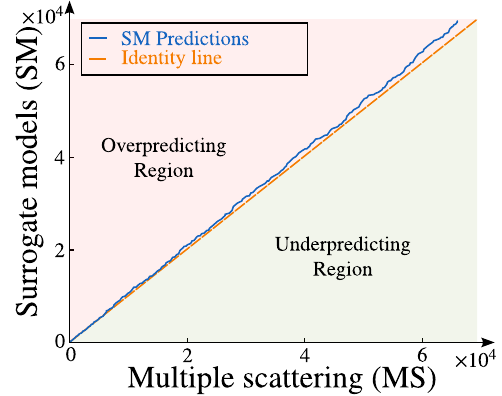}
\caption{Power per unit volume error.}
\label{fig:error_one2one}
\end{subfigure}
\captionsetup[figure]{justification=centering}
\centering
\caption{Characterization of the performance of surrogate models in estimating the objective function using $20{,}000$ randomly selected specifications (plant, control, and layout) for a $5$-WEC farm.} 
\label{fig:obj_error_validatoin}
\end{figure}

The error introduced by estimating the hydrodynamic coefficients through surrogate models may propagate and adversely impact the estimation of power and, thus, the objective function.
To gain a better understanding of how different design specifications affect the power estimation, $20{,}000$ samples from the entire design space, including plant ([$\Rwec$,$\RDwec$]), control ([$\BPTO$, $\KPTO$]), and layout ($\AL$) were randomly selected for a $5$-WEC farm.
To ensure compatibility with the optimization problem, a minimum safety distance of $10~[\unit{m}]$ was included in the simulation to allow maintenance ships to pass safely.
The analysis was carried out using the SM, as well as the MS approach.
The results, which are shown in Figs.~\ref{fig:error_abs_obj} and \ref{fig:error_one2one}, indicate that the absolute error at the $99$th percentile of samples is $0.38$ [$\unit{MW/m^3}$].
While extreme cases with a higher magnitude of error are possible, the probability of their occurrence is relatively small.
In addition, Fig.~\ref{fig:error_one2one} points to the relatively close and thus acceptable performance of the surrogate models.
While the objective function values predicted through surrogate models follow the identity line relatively closely, there is a small level of overprediction, particularly in higher values of the objective function (blue line in Fig.~\ref{fig:error_one2one}). 

With the first two methods for SM validation shown, we can now conclude that the resulting surrogate models offer reasonable accuracy in predicting the hydrodynamic coefficients, as well as the power per unit volume objective function.
Despite these outcomes, we still need to ensure that our models are capable of identifying the optimal design of the farm.
This task is addressed through case studies presented in the next section.

\section{Results and Discussion}
\label{sec:Optimization}

In this section, we first discuss the concurrent geometry, control, and layout optimization formulation and then introduce several case studies to assess the performance of SMs within an optimization problem.
This section includes a series of problems with an increasing level of complexity, concluding with an integrated plant geometry, control, and layout optimization problem of a $25$-WEC farm.

For the realization of a balanced, system-level design framework, a necessary step is to identify an objective function that represents the intricacies of plant, control, and layout disciplines.
The idea of only maximizing power within this system-level framework often leads to WEC devices that are too large to be economically viable. 
While economic metrics such as net present value and levelized cost of energy are suitable for such investigations, there is currently limited information regarding how such costs scale with variations in the size of the device, the control effort, and the size of the farm.
As shown in Eq.~(\ref{eq:powerpervolumeObj}), a straightforward remedy is to maximize power per unit volume of the device \citep{falnes2020ocean}.
While this approach normalizes power by the volume of the device (i.e., generally leading to smaller WECs), the presence of a poorly-posed control optimization problem creates additional limitations.
Specifically, the control optimization problem is not well-posed in the frequency domain because it results in impractically-large device motions and, thus, unrealistic power outputs. 
While it is possible to address this issue through advanced control strategies, such as pseudo-spectral controllers \citep{Coe2020} or model predictive control \citep{sergiienko2021effect}, such investigations are beyond the scope of the current study.

To mitigate some of these limitations, we directly impose various power saturation limits $P_\text{lim}$ on the WEC power matrix $\mathbf{p}_{i}(\cdot)$ of Eq.~(\ref{eq:pto_power3}) to directly restrain the power that can be produced through the device.
The implication is that the power produced for some of the sea states, and thus the objective function, may be flat for a certain range of PTO parameters. 
A practical implementation would include costs associated with the control effort to avoid this challenge.
When no power saturation is applied, the maximizing power strategy is the one that maximizes the motions of the WEC above practical limits.
While this control strategy results in large motions and forces with unrealistic performance values within the optimization problem, it is often insightful to better understand the limits of device performance.

With these limitations in mind, we first introduce the concurrent plant geometry, control, and layout optimization formulation and then present several case studies with an increasing level of integration.
Case Studies \rom{1} and \rom{2} validate the performance of surrogate models within an optimization problem, focusing on geometry and layout optimization, respectively.
The latter also discusses the usage of a hybrid optimization approach to improve upon the optimized layout solution obtained through SM.
Case Studies \rom{3}, \rom{4}, and \rom{5} are devised to increase the level of integration within the problem by concurrently considering the optimization of design domains associated with geometry, control, and layout.
Case Study \rom{6} assesses the performance of the proposed approach for a $25$-WEC farm to address scalability questions. 
These case studies are briefly described below:
\begin{enumerate}[topsep=0pt,itemsep=-1ex,partopsep=1ex,parsep=1ex,label=$\bullet$]
    \item Case Study \rom{1}:~Geometric plant optimization of a $5$-WEC farm (Sec.~\ref{subsubsec:CaseStudy1})
    \item Case Study \rom{2}:~Layout optimization of a $5$-WEC farm (Sec.~\ref{subsubsec:CaseStudy2})
    \item Case Study \rom{3}:~Concurrent geometric plant and layout optimization of a $5$-WEC farm (Sec.~\ref{subsubsec:CaseStudy3})
    \item Case Study \rom{4}:~Concurrent geometric plant, farm-level control, and layout optimization of a $5$-WEC farm (Sec.~\ref{subsubsec:CaseStudy4})
    \item Case Study \rom{5}:~Concurrent geometric plant, device-level control, and layout optimization of a $5$-WEC farm (Sec.~\ref{subsubsec:CaseStudy5})
    \item Case Study \rom{6}:~Concurrent geometric plant, farm-level control, and layout optimization of a $25$-WEC farm (Sec.~\ref{subsubsec:CaseStudy6})
\end{enumerate}

\subsection{Problem Formulation}
\label{subsec:Problem_Formulation}

\begin{table}[b]
    \caption{Problem parameters.}
    \label{Tab:Parameter_Opt}
    \centering
    \begin{tabular}{s c s c}
    \hline  \hline
    \textrm{\textbf{Option}} & \textrm{\textbf{Value}} &  \textrm{\textbf{Option}} & \textrm{\textbf{Value}} \\
    \hline
    $\Rwecmin$ & $0.5~[\unit{m}]$ & $\Rwecmax$ & $10~[\unit{m}]$ \\
    $\RDwecmin$ & $0.2$ & $\RDwecmax$ & $10$ \\
    $\Dwecmin$ & $0.5~[\unit{m}]$ & $\Dwecmax$ & $20~[\unit{m}]$ \\
    $\KPTOmin$ & $-\KPTOmax$ & $\KPTOmax$ & $5\times 10^{5}~[\unit{N/m}]$ \\
    $\BPTOmin$ & $0~ [\unit{Ns/m}]$ & $\BPTOmax$ & $5\times 10^{5}~[\unit{Ns/m}]$\\
    $\underaccent{\bar}{\bm{x}}$ & $0~[\unit{m}]$ &  $\bar{\bm{x}}$ & $0.5\sqrt{2\Nwec \times 10^4}~[\unit{m}]$ \\
    $\underaccent{\bar}{\bm{y}}$ & $-\bar{\bm{y}}$ &  $\bar{\bm{y}}$ & $0.5\sqrt{2\Nwec \times 10^4}~[\unit{m}]$ \\
    $\rho$ & $1025~ [\unit{kg/m^3}]$ & $g$ & $9.81~[\unit{m/s^{2}}]$ \\
    $s_{d}$ & $10~[\unit{m}]$ & $\Nwec$ & $5$ \\
    $n_{yr}$   & $30~[\unit{years}]$ & $n_{r}$  & $200$\\
    $n_{gq}$   & $500$  & $\eta_{\text{pcc}}$ & $0.8$ \\
    $\eta_{\text{oa}}$ & $0.95$ & $\eta_{\text{t}}$ & $0.98$\\
    \hline \hline
    \end{tabular}
\end{table}

Here, we present the general formulation for a concurrent plant, control, and layout optimization of WEC farms. 
Since manufacturing costs associated with WECs of different dimensions become prohibitive, it is reasonable to assume that WEC dimensions are uniform across the farm.
Depending on the case study, the control parameters can be optimized at the farm level (uniform across the farm $\bm{u} \in \mathbb{R}^{{2}}$) or device level ($\bm{u} \in \mathbb{R}^{{2}\Nwec}$).
Finally, while some studies impose symmetric constraints \citep{lyu2019optimization}, this approach is avoided in this study.
Expressing the objective function as the average power per unit volume of the device, the concurrent geometry, control, and layout optimization problem can be formulated as:
\begin{subequations}
 \label{Eqn:OPtimization}
 \begin{align}
 \underset{\bm{p},\bm{u}, \AL}{\textrm{minimize:}}
 \quad & - p_{v}(\bm{p}, \bm{u}, \AL)   \label{Eqn:Obj} \\
 \textrm{subject to:} \quad
    \begin{split}
    &  2\Rwec + s_{d} - \bm{L}_{pq} \leq 0 \quad\\
         & \qquad \forall ~~  p,q = 1, 2, \dots, \Nwec \quad p \neq q \label{Eqn:distanceconst}
    \end{split} \\
    & \mathbf{p}_{i}(\cdot) \leq p_{\text{limit}}~\text{(optional)} \label{Eqn:PowerSaturation} \\
    & \Dwecmin \leq \Dwec \leq \Dwecmax \label{Eqn:draftcons} \\
    &  \underaccent{\bar}{\bm{p}} \leq \bm{p} \leq \bar{\bm{p}}\label{Eqn:plantconst}\\
    &  \underaccent{\bar}{\bm{u}} \leq \bm{u} \leq \bar{\bm{u}} \label{Eqn:controlconst}\\
    &  \underaccent{\bar}{\AL} \leq \AL \leq \bar{\AL}\label{Eqn:layoutconst}\\
    \textrm{where:} \quad & \bm{p} = [\Rwec, \RDwec]^{T} \in \mathbb{R}^{2} \notag\\
                 & \bm{u} = [\KPTO, \BPTO]^{T} \in \mathbb{R}^{{2}} ~~\text{or}~~ \mathbb{R}^{{2}\Nwec}  \notag \\
                & \AL = [\bm{x},\bm{y}] \in \mathbb{R}^{{2}(\Nwec-1)} \notag \\
                & \Dwec = \Rwec/\RDwec \notag 
 \end{align} 
\end{subequations}
\noindent
where $\bm{p}$ is the vector of geometric plant optimization variables composed of WEC radius $\Rwec$ and slenderness ratio $\RDwec$ (uniform across the farm). 
$\bm{u}$ is the vector of time-independent control parameters of the WEC devices, composed of PTO spring stiffness $\KPTO$ and damping $\BPTO$.
The layout design vector $\AL$ is composed of $(\Nwec - 1)$ coordinates, corresponding to the $x$- and $y$-axis locations of WEC devices.
Note that here, the first WEC is always assumed to be positioned at the center of the coordinate system and, thus, is not included in $\AL$. 
Equation~(\ref{Eqn:distanceconst}) ensures that the minimum distance between each pair of WECs is larger than the WEC diameter plus an additional safe distance $s_{d}$, which is required to allow maintenance ships to pass safely \citep{neshat2022layout}.
Here, a safety distance of $s_{d} = 10~[\unit{m}]$ is consistently used.
Note that Eq.~(\ref{Eqn:distanceconst}) is present only for optimization problems involving layout. 
When included, Eq.~(\ref{Eqn:PowerSaturation}) imposes a power saturation limit on the device power matrix.
Equation~(\ref{Eqn:draftcons}) ensures that the draft of WEC devices is practically viable.
All of the optimization variables are confined within their associated bounds, as described by Eqs.~(\ref{Eqn:plantconst})--(\ref{Eqn:layoutconst}).
The farm area is restricted to a box with the width of $0.5 \times\sqrt{20000\Nwec}~\textrm{m}$ in $x$ and  $\pm 0.5 \times\sqrt{20000\Nwec}~\textrm{m}$ in $y$ axis \citep{neshat2022layout}.

These details, along with some additional problem data, are presented in Table~\ref{Tab:Parameter_Opt}.
From the table, it is clear that we have allowed the PTO stiffness to take negative values.
This enables a reactive phase control strategy \citep{antonio2010wave} that is often used to extend the range of resonance conditions in WEC design \citep{todalshaug2016tank, peretta2015effect}.
Despite the limitations of this approach, as stated in \cite{antonio2010wave}, it is shown
that phase control with a negative spring stiffness increases the
absorbed power \citep{todalshaug2016tank}.

\subsection{Results and Discussion}
\label{subsec:Results_and_Discussion}

So far, we have validated our SMs and concluded that they offer reasonable accuracy in estimating the hydrodynamic coefficients, as well as the power captured per unit volume of the device.
A remaining task is to ensure that SMs are effective in finding the optimal solution. 
Therefore, before introducing case studies with higher levels of integration, we will assess the performance of SMs in the context of geometric plant and layout optimization problems in Case Study \rom{1} and \rom{2}, respectively.

\subsubsection{Case Study \rom{1}: Geometric Plant Optimization}
\label{subsubsec:CaseStudy1}

WEC geometry has a considerable impact on the cost and performance of the device.
As discussed in \cite{Coe2021}, WECs with larger dimensions are often capable of harvesting larger amounts of power at a smaller range of frequencies.
Smaller WECs, on the other hand, generally capture lower levels of power at a higher range of frequencies.
This conclusion directly ties the design of WEC geometry to the site location and its environmental and climate attributes.
The WEC geometry design, nevertheless, is integrated and thus affected by the WEC control and layout. 

\begin{figure}[t]
\captionsetup[subfigure]{justification=centering}
\centering
\begin{subfigure}{0.5\columnwidth}
\centering
 \includegraphics[scale=0.49]{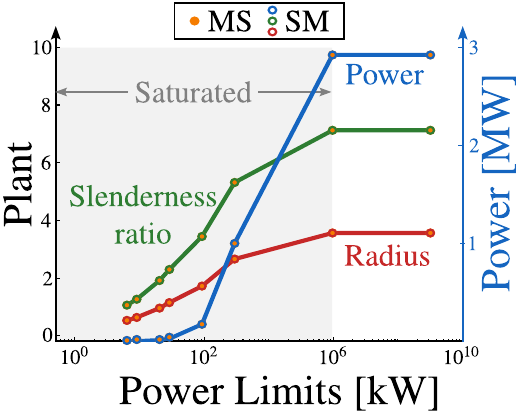}
\caption{East Coast (site id: NAEC8).}
\label{fig:P_opt_east}
\end{subfigure}%
\begin{subfigure}{0.5\columnwidth}
\centering
\includegraphics[scale=0.49]{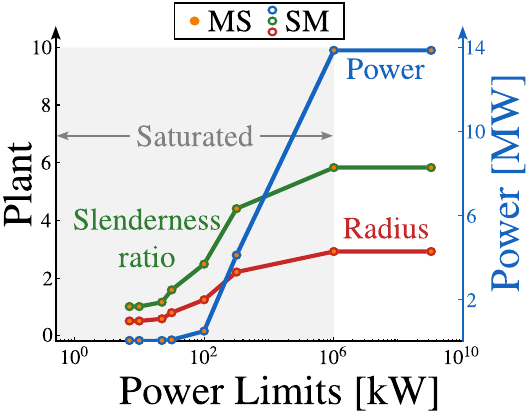}
\caption{West Coast (site id: N46229).}
\label{fig:P_opt_west}
\end{subfigure}
\captionsetup[figure]{justification=centering}
\centering
\caption{Case Study \rom{1}: Geometric plant optimization study for a single WEC located on (a) East Coast, and (b) West Coast of the United States using \texttt{Matlab}'s \texttt{fmincon}. Optimal radius, slenderness ratio, and power are plotted for various power saturation limits using both surrogate modeling (SM) and multiple scattering (MS) with
prescribed control parameters $\BPTO = 500~[\unit{kNs/m}]$ and $\KPTO = -0.5~[\unit{kN/m}]$.} 
\label{fig:P_opt}
\end{figure}

Overlooking the coupling with control and layout domains, this simplified case study provides additional evidence that the proposed method and the resulting surrogate models are effective in efficiently solving the optimization problem.
This study is carried out for various power saturation limits using SM in combination with MBE for a single WEC device located on the East Coast and West Coast.
Using \texttt{Matlab}'s \texttt{fmincon} with an interior-point algorithm, the optimized solutions using SM are compared with the solution obtained using MS, and the results are presented in Table~\ref{tab:p_opt}, and Fig.~\ref{fig:P_opt}.

\begin{table}[t]
\centering
\caption{Case Study \rom{1}: Geometric plant optimization study with prescribed PTO parameters and different power saturation limits for a single WEC device located on the East Coast and West Coast of the United States, obtained using \texttt{Matlab}'s \texttt{fmincon}. The hydrodynamic coefficients were calculated using both surrogate modeling (SM) and multiple scattering (MS) with prescribed $\BPTO = 500~[\unit{kNs/m}]$ and $\KPTO = -0.5~[\unit{kN/m}]$.}
\label{tab:p_opt}

\begin{threeparttable}

\makebox[\linewidth]{%
\begin{tabular}{c c r r r r r}
\hline \hline
\multirow{2}{*}{\rotatebox[origin=c]{0}{\textbf{Site}}} & \multirow{2}{*}{\rotatebox[origin=c]{0}{$p_{\text{lim}}\tnote{a}$}} & \multirow{2}{*}{\rotatebox[origin=c]{0}{\textbf{Method}}}  & \multicolumn{2}{c}{\textbf{Plant}} & \multirow{2}{*}{\textrm{{P}}\tnote{c}}\\
\cline{4-5} 
& & & $\Rwec~[\unit{m}]$ & $\RDwec$ \\
\hline 
\multirow{10}{*}{\rotatebox[origin=c]{90}{East Coast}} & \multirow{2}{*}{$10^0$} & \text{MS} & $0.6318$ & $1.2631$ & $0.0027$ \\
&  & \text{SM} & $0.6329$ & $1.2652$ & $0.0027$ \\ \cline{3-6}
&  \multirow{2}{*}{$10^1$} & \text{MS}  &  $1.1477$ & $2.2950$ & $0.0273$ \\
&  & \text{SM}  &  $1.1468$ & $2.2933$ & $0.0272$ \\ \cline{3-6}
&  \multirow{2}{*}{$10^2$} & \text{MS}  &  $1.7204$ & $3.4408$ & $0.1753$ \\
&  & \text{SM}  &  $1.7224$ & $3.4441$ & $0.1757$ \\ \cline{3-6}
&  \multirow{2}{*}{$10^3$} & \text{MS}  &  $2.6571$ & $5.3138$ & $1.0001$ \\
&  & \text{SM}  &  $2.6593$ & $5.3182$ & $1.0017$ \\ \cline{3-6}
&  \multirow{2}{*}{$10^6$} & \text{MS}  &  $3.5618$ & $7.1236$ & $2.9228$ \\
&  & \text{SM}  &  $3.5681$ & $7.1362$ & $2.9331$ \\
\hline
\multirow{10}{*}{\rotatebox[origin=c]{90}{West Coast}} & \multirow{2}{*}{$10^0$} & \text{MS} & $0.5000$ & $1.0000$ & $0.0077$ \\
&  & \text{SM} & $0.5000$ & $1.0000$ & $0.0077$ \\ \cline{3-6}
&  \multirow{2}{*}{$10^1$} & \text{MS}  &  $0.7881$ & $1.5761$ & $0.0635$ \\
&  & \text{SM}  &  $0.7885$ & $1.5768$ & $0.0636$ \\ \cline{3-6}
&  \multirow{2}{*}{$10^2$} & \text{MS}  &  $1.2346$ & $2.4693$ & $0.4805$ \\
&  & \text{SM}  &  $1.2350$ & $2.4700$ & $0.4807$ \\ \cline{3-6}
&  \multirow{2}{*}{$10^3$} & \text{MS}  &  $2.1946$ & $4.3893$ & $4.0904$ \\
&  & \text{SM}  &  $2.1965$ & $4.3931$ & $4.0973$ \\ \cline{3-6}
&  \multirow{2}{*}{$10^6$} & \text{MS}  &  $2.9052$ & $5.8105$ & $13.8206$ \\
&  & \text{SM}  &  $2.9108$ & $5.8216$ & $13.8734$ \\ 
\hline \hline
\end{tabular}
}

\begin{tablenotes} [para,flushleft] \footnotesize
\item [a] Power saturation limit prescribed in [\unit{kW}]
\item [c] Power over the lifetime of the device calculated in [\unit{MW}]
\end{tablenotes}

\end{threeparttable}
\end{table}

According to Table~\ref{tab:p_opt}, for both locations, the optimized plant, as well as generated power over the lifetime of the device using surrogate modeling (in combination with MBE), closely match the solution obtained using MS.
On average, the solution obtained through SM signifies a $98.67\%$ increase in computational efficiency compared to MS.
While providing additional evidence for the acceptable performance of SMs, this investigation also points to the coupling between plant and control domains.
Specifically, as shown in Fig.~\ref{fig:P_opt}, optimal WEC radius and slenderness ratio change with variations in power saturation limit (which can be considered a control parameter).
According to these plots, in both locations, the optimal plant variables increase when we increase power saturation limits, until they plateau for cases in which the power saturation limit exceeds the available wave resources in the region. 
The coupling between plant and control demonstrated here is further corroborated in \cite{GarciaRosa2016} for latching, declutching, and model predictive control technologies.

The impact of site location on plant optimality is also evident from these results.
For the same controller design, the WEC device on the East Coast is larger in radius and slenderness ratio.
Therefore, the WEC designed for the East Coast is capable of capturing higher amounts of power at a smaller range of frequencies compared to the WEC designed for the West Coast.

In all of the results presented in Table~\ref{tab:p_opt}, the WEC draft dimension remains at its lower bound of $\Dwec = 0.5~[\unit{m}]$.
As presented, the WEC model entails certain assumptions that directly affect WEC draft value.
Specifically, currently, there is no mechanism to prevent the WEC from leaving the water.
In addition, using the unit per volume power objective function might contribute to smaller draft size.
These observations point to some of the limitations in the current model.
Nevertheless, under the current assumptions, the draft dimension seems to be independent of radius, PTO control, and layout for the geometric plant optimization study.

\subsubsection{Case Study \rom{2}: Layout Optimization}
\label{subsubsec:CaseStudy2}

\begin{figure*}[t]
\captionsetup[subfigure]{justification=centering}
\centering
\begin{subfigure}{\textwidth}
\centering
\includegraphics[scale=\xywecscale]{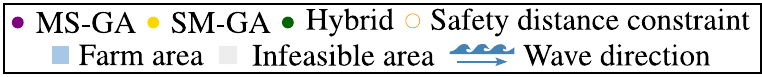}
\label{subfig:Legend_L_opt}
\end{subfigure}
\begin{subfigure}{0.25\textwidth}
\centering
\includegraphics[scale=\xywecscale]{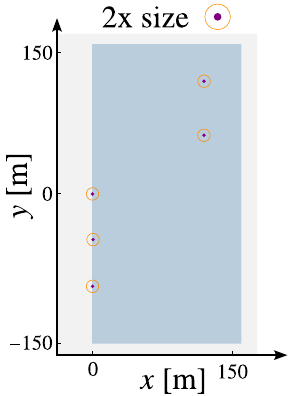}
\caption{MS.}
\label{fig:L_opt_MS}
\end{subfigure}%
\begin{subfigure}{0.25\textwidth}
\centering
\includegraphics[scale=\xywecscale]{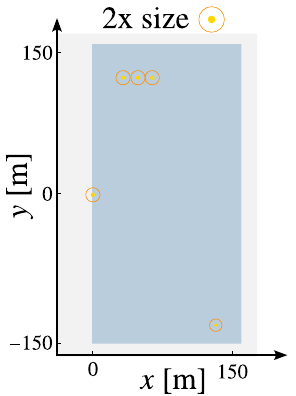}
\caption{SM.}
\label{fig:L_opt_SM}
\end{subfigure}%
\begin{subfigure}{0.25\textwidth}
\centering
\includegraphics[scale=\xywecscale]{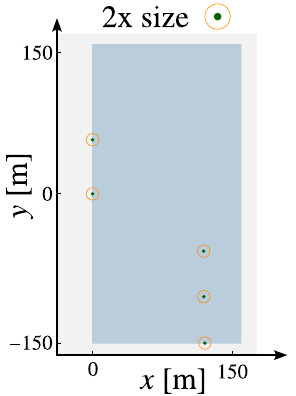}
\caption{Hybrid.}
\label{fig:L_opt_Hybrid}
\end{subfigure}
\captionsetup[figure]{justification=centering}
\centering
\caption{Case Study \rom{2}: Layout optimization study for a $5$-WEC farm with $\Rwec = 2~[\unit{m}]$, $\RDwec = 1$, $\BPTO = 500~[\unit{kNs/m}]$, and $\KPTO = -5~[\unit{kN/m}]$ located on the West Coast.
{\texttt{MATLAB}'s \texttt{ga}} is used for multiple scattering (MS) and surrogate modeling (SM), while both {\texttt{ga}} and \texttt{fmincon} are used for the hybrid method.
The WECs and their associated constraints are drawn in scale.
Depiction of a single WEC with $2$x scaling is presented at the top of each plot for improved visualization. } 
\label{fig:L_opt}
\end{figure*}

Due to its complexity and the presence of multiple local minima, the layout optimization study requires a global optimization approach, such as a genetic algorithm (GA).
Solving a layout optimization problem using GA with MS, however, is computationally prohibitive.
For example, for a $5$-WEC farm, the layout optimization problem takes approximately $42.6$ hours per GA generation, or $26.6$ days for the problem to converge.
Therefore, it is critical to identify and implement computationally efficient approaches to identify superior layout configurations.
For a given site located on the West Coast with prescribed radius $\Rwec = 2~[\unit{m}]$, slenderness ratio $\RDwec = 1$, and control parameters $\BPTO = 500~[\unit{kNs/m}]$, and $\KPTO = -5~[\unit{kN/m}]$, with no power saturation limit, the layout optimization problem is first carried out using MS.
Next, the problem is solved using the proposed framework of SM.
The results from this investigation are presented in Table~\ref{Tab:l_opt} and Fig.~\ref{fig:L_opt}. 

\begin{table}[t]
\centering
\caption{Case Study \rom{2}: Layout optimization study for a $5$-WEC farm with prescribed plant $\Rwec = 2~[\unit{m}]$, $\RDwec = 1$, and control $\BPTO = 500~[\unit{kNs/m}]$, $\KPTO = -5~[\unit{kN/m}]$. {\texttt{MATLAB}'s \texttt{ga}} is used for multiple scattering (MS) and surrogate modeling (SM), while both {\texttt{ga}} and \texttt{fmincon} are used for the hybrid method.}
\label{Tab:l_opt}

\begin{threeparttable}
  
\makebox[\linewidth]{%
\begin{tabular}{r r r r r r}
\hline  \hline
\multirow{1}{*}{\textrm{\textbf{Method}}}  & \multirow{1}{*}{$L$\tnote{a}} & \multirow{1}{*}{\textrm{P}\tnote{b}} & $q_\text{factor}$ & \multirow{1}{*}{\textrm{\textbf{Time}}\tnote{c}}   \\
\hline 
\multirow{1}{*}{\textrm{MS-GA}\tnote{d}} & \multirow{1}{*}{\textrm{Fig.~\ref{fig:L_opt_MS}}}  &  $25.51$ & $1.0044$ & $639.32$ \\
\multirow{1}{*}{\textrm{SM-GA}\tnote{e}} & \multirow{1}{*}{\textrm{Fig.~\ref{fig:L_opt_SM}}} & $25.27$  & $0.9952$  & $0.90$ \\
\multirow{1}{*}{\textrm{Hybrid}\tnote{f}} &\multirow{1}{*}{\textrm{Fig.~\ref{fig:L_opt_Hybrid}}} & $25.51$  & $1.0044$ & $6.02$ \\
\hline \hline
\end{tabular}
}
    
\begin{tablenotes} [para,flushleft] \footnotesize
\item [a] Optimized layout%
\item [b] Power over farm's lifetime using MS in [\unit{MW}]%
\item [c] Computational time in [\unit{h}]%
\item [d] Multiple scattering with {\texttt{ga}}%
\item [e] Surrogate modeling with {\texttt{ga}}%
\item [f] Hybrid optimization approach consisting of {SM with \texttt{ga} + MS with \texttt{fmincon}}%
\end{tablenotes}

\end{threeparttable}
\end{table}

According to Table~\ref{Tab:l_opt}, the optimized farm power using SM results in approximately $0.94\%$ error compared to the MS solution.
While this error is relatively small, further investigation of the optimized farm layout points to the potential for additional improvements.
Specifically, according to Fig.~\ref{fig:L_opt_SM}, the optimized layout using SM results in some shadowing effect that can reduce the performance of the farm.
This is evidenced by the fact that the q factor $q_\text{factor} = 0.9952$ is less than $1$, indicating that negative interactions are present in the optimized layout solution.
It is evident that while the proposed method is effective in finding the optimal plant (as evidenced by Sec.~\ref{subsubsec:CaseStudy1}), it can not always find the optimal layout, particularly when the contributions from the farm layout are small (for instance, due to a poor controller design).

An effective strategy to address this issue is to utilize a hybrid optimization problem, in which the GA solution using SM is a starting point for a gradient-based optimization solver that utilizes MS to improve upon the optimized layout.
This hybrid approach, while expensive compared to SM, is still more computationally efficient than using MS with GA, offering a $91$-fold increase in computational efficiency. 
Using the hybrid optimization approach, the optimized farm power is estimated to be $25.51$, which indicates a $0.00006\%$ difference in power estimation in comparison to the MS solution, and an increase in the q factor $q_\text{factor} = 1.0044$. 
This result indicates that using a hybrid optimization approach is an effective way of finding optimal layouts more efficiently than the direct use of MS.

Optimized layout configurations associated with this investigation are presented in Fig.~\ref{fig:L_opt}.
The optimized layout obtained using SM, shown in Fig.~\ref{fig:L_opt_SM} is improved by implementing a hybrid optimization approach in Fig.~\ref{fig:L_opt_Hybrid}.
Comparing the layout solutions obtained using MS and the hybrid optimization approach, it is clear that the layouts are very similar, with two clusters of $2$ and $3$ WEC devices in two rows. 
This observation points to the presence of multiple layout configurations with comparable performances across the farm. 

\begin{figure}
\captionsetup[subfigure]{justification=centering}
\centering
\begin{subfigure}{0.5\columnwidth}
\centering
\includegraphics[scale=\twofigscale]{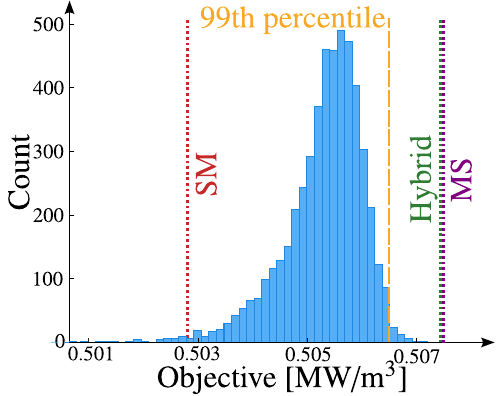}
\caption{}
\label{subfig:L_opt_Analysis_1}
\end{subfigure}%
\begin{subfigure}{0.5\columnwidth}
\centering
\includegraphics[scale=\twofigscale]{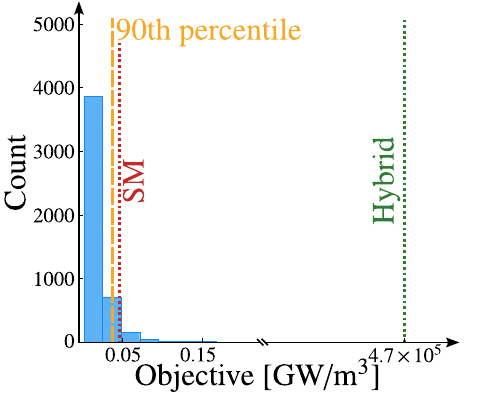}
\caption{}
\label{subfig:L_opt_Analysis_2}
\end{subfigure}
\captionsetup[figure]{justification=centering}
\centering
\caption{Performance comparison of optimized WEC layouts to $5{,}000$ randomly-selected layout configurations with similar plant and control specifications. Solutions from multi scattering (MS), surrogate modeling (SM), and hybrid approaches are also plotted: (a) Prescribed plant and control parameters are not system-level optimal and are associated with Table~\ref{Tab:l_opt}, (b) Prescribed plant and device-level control parameters are based on a system-level optimal solution from Table~\ref{Tab:PCL_device_None}.}
\label{fig:L_opt_Analysis}
\end{figure}

The results presented in Table~\ref{Tab:l_opt} correspond to an investigation in which the impact of layout is relatively small.
This observation is further illustrated by comparing the objective function of $5{,}000$ randomly-selected layout configurations with similar plant and control specifications that satisfy all problem constraints to the optimal solutions presented in Table~\ref{Tab:l_opt}. 
The results from this analysis are shown in Fig.~\ref{subfig:L_opt_Analysis_1}, indicating that for a variety of layouts, the objective function changes within the range of $0.006~[\unit{MW/m^{3}}]$. 
In addition, it is observed that the hybrid and MS solutions outperform the $99\%$ percentile of the randomly-selected layouts.
The small range of objective function variation in Fig.~\ref{subfig:L_opt_Analysis_1} is associated with the fact that the impact of plant and controller design on optimal farm layout is overlooked.
When concurrent plant, control, and layout are considered, however, the objective function changes within a much broader range.
This outcome is shown in Fig.~\ref{subfig:L_opt_Analysis_2}, where the optimal plant and device-level control solutions (from Sec.~\ref{subsubsec:CaseStudy5}) are prescribed with no power saturation limits and farm simulations were implemented for $5{,}000$ randomly-generated layouts that satisfy all problem constraints.
While this observation may change by considering more realistic and practical control limits, it highlights the coupling between the layout and controller design of WEC farms.
From Fig.~\ref{subfig:L_opt_Analysis_2}, it is clear that the SM solution outperforms $90$th percentiles of the simulated farm layouts.
In addition, the hybrid approach enables a significantly better solution by fine-tuning plant, control, and layout decision variables.
While more practical controller designs can affect this coupling, the analysis presented in Fig.~\ref{fig:L_opt_Analysis} highlights both the role of layout optimization and its integration on farm performance, as well as the effectiveness of SM and the hybrid optimization approach in finding superior layouts.

A remaining question is to assess the proximity of the optimized layout solution from SM to the solution from MS.
To answer this question, we perform a sensitivity study of the SM-optimized layout solution using MS.
Specifically, after prescribing the plant and control parameters, we fix all the WEC devices at their optimized location, except for the WEC under investigation, which is perturbed using a large number of random simulations within a specified radius of its optimized location and randomly-selected angles.   
The objective function is then obtained using MS, and its contours are plotted to provide insight into the objective function value in the neighborhood of the optimized solution.  
Next, a hybrid layout optimization problem was carried out using \texttt{fmincon} and MS, in which all WEC devices were fixed at their optimized locations, except for the WEC under investigation.
Fixing all but the WEC of interest is necessary for the sake of this validation effort in order to limit the interactions caused by the movement of other WEC devices.
Therefore, the assumption of fixing all other WECs at their optimized location is only necessary in this analysis for the sake of comparison.
All of the other hybrid optimization solutions discussed in the article allow the layout and additional decision variables to vary by the optimizer.

\begin{figure}
\captionsetup[subfigure]{justification=centering}
\centering
\begin{subfigure}{\columnwidth}
\centering
\includegraphics[scale=\twofigscale]{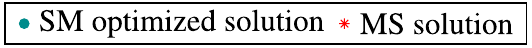}
\label{subfig:L_opt_Analysis2_leg}
\end{subfigure}
\begin{subfigure}{0.5\columnwidth}
\centering
\includegraphics[scale=\twofigscale]{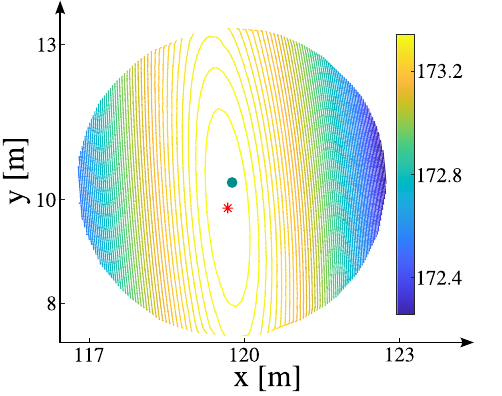}
\caption{WEC \# 2.}
\label{subfig:L_opt_Analysis2_WEC2}
\end{subfigure}%
\begin{subfigure}{0.5\columnwidth}
\centering
\includegraphics[scale=\twofigscale]{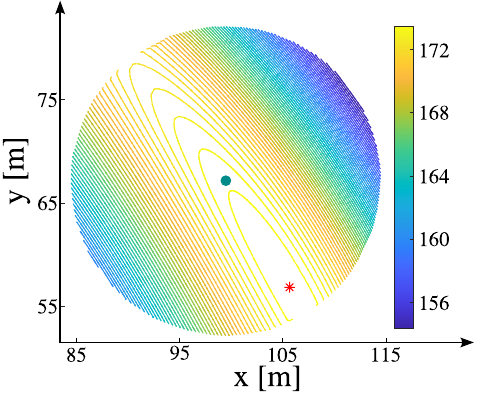}
\caption{WEC \# 3.}
\label{subfig:L_opt_Analysis2_WEC3}
\end{subfigure}
\captionsetup[figure]{justification=centering}
\centering
\caption{Sensitivity analysis of the optimized farm layout obtained using surrogate modeling (SM). The objective function contours are created in $[\unit{MW/m^3}]$ using a multiple scattering (MS) approach, with a large number of random simulations in the neighborhood of the optimized SM solutions. The hybrid solution is obtained by keeping all other WECs at their optimized locations and only allowing the optimizer to move the WEC of interest.}
\label{fig:L_opt_Analysis2}
\end{figure}

Using optimized results from an investigation on the Alaska Coast, the sensitivity analysis is carried out, and the results for two of the WECs (WEC \# 2 and WEC \# 3) within the farm are visualized in Fig.~\ref{fig:L_opt_Analysis2}, using the polar contour plots developed by \cite{MFX38858}.
From Fig.~\ref{subfig:L_opt_Analysis2_WEC2}, it is clear that for WEC \#2, the optimized solution obtained using SM is only $0.5~[\unit{m}]$ away from the neighboring local optimum.
The difference in the objective value in this case is $0.95\%$.
For WEC \# 3, however, the optimized location obtained using SM is $12.02~[\unit{m}]$ away from the optimal hybrid solution.
This distance is shown in Fig.~\ref{subfig:L_opt_Analysis2_WEC3}.
Note that while this distance seems too far, the optimized objective from SM only differs from the hybrid solution by $1.18\%$.
Therefore, the performance of the surrogate models is acceptable.
Nevertheless, there is certainly potential for improvement, particularly by accounting for and quantifying the inherent epistemic uncertainty in these models for future work.
For now, as shown through numerous validation efforts, using SM in combination with a hybrid approach offers a promising solution to deal with inherent errors in surrogate modeling.

\subsubsection{Case Study \rom{3}: Geometric Plant and Layout Optimization}
\label{subsubsec:CaseStudy3}

Assuming that the PTO control parameters are prescribed, this case study investigates the concurrent geometric plant and layout optimization of a $5$-WEC farm for all four locations on the Alaska Coast, East Coast, Pacific Islands, and West Coast.
To further signify the impact of control parameters, an additional study with power saturation limits is also included in this section.

Prescribing the PTO parameters as $\BPTO = 500~[\unit{kNs/m}]$, and $\KPTO = -5~[\unit{kN/m}]$ with no power saturation limits, the concurrent plant and layout optimization problem is formulated and solved for a $5$-WEC farm at various locations, and the results are tabulated in Table~\ref{Tab:LP_opt_Plim_none}. 
According to this table, the radius of the WEC device changes in the range of $2.91~[\unit{m}]$ for the West Coast (N46229) and $3.59~[\unit{m}]$ for the East Coast (NAEC8).
With the exception of the design for the Alaska Coast (NAWC24), the WEC radius changes inversely with the power capture of the farm.
The slenderness ratio of the WEC device also changes for each location; however, the draft dimension remains at its lowest bound of $0.5~[\unit{m}]$ for all cases.
As discussed before, this seems to be related to multiple factors including the choice of the objective function, as well as the absence of practical and realistic constraints to limit the motion of the WEC device according to its draft dimension.
The latter, which is directly related to the usage of frequency-domain models in this study, points to one of the limitations and future directions of the work. 

\begin{figure*}[t]
    \captionsetup[subfigure]{justification=centering}
    \centering
    \begin{subfigure}{\columnwidth}
    \centering
    \includegraphics[scale=\xywecscale]{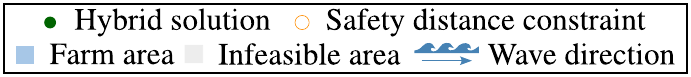}
    \label{subfig:LP_opt_Legend_1}
    \end{subfigure}
    \begin{subfigure}{0.25\textwidth}
    \centering
    \includegraphics[scale=\xywecscale]{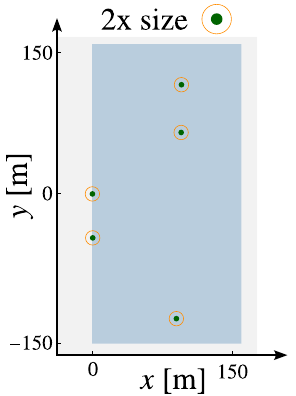}
    \caption{Alaska Coast (NAWC24).}
    \label{subfig:LP_opt_none_Alaska}
    \end{subfigure}%
    \begin{subfigure}{0.25\textwidth}
    \centering
    \includegraphics[scale=\xywecscale]{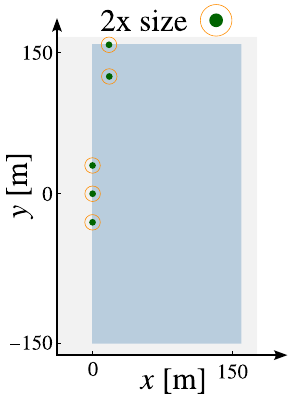}
    \caption{East Coast (NAEC8).}
    \label{subfig:LP_opt_none_East}
    \end{subfigure}%
    \begin{subfigure}{0.25\textwidth}
    \centering
    \includegraphics[scale=\xywecscale]{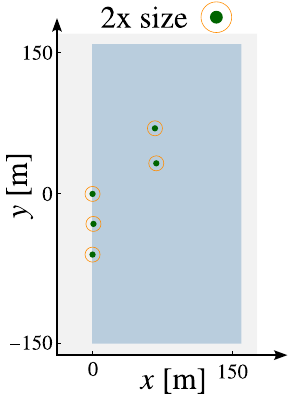}
    \caption{Pacific Islands (PI14).}
    \label{subfig:LP_opt_none_Pacific}
    \end{subfigure}%
    \begin{subfigure}{0.25\textwidth}
    \centering
    \includegraphics[scale=\xywecscale]{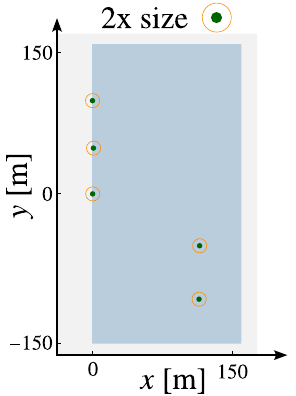}
    \caption{West Coast (N46229).}
    \label{subfig:LP_opt_none_West}
    \end{subfigure}%
    \captionsetup[figure]{justification=centering}
    \caption{Case Study \rom{3}: Optimized layout from the concurrent geometric plant and layout optimization study for a $5$-WEC farm at different site locations using the hybrid optimization approach {(SM with \texttt{ga} + MS with \texttt{fmincon})} using $\BPTO = 500~[\unit{kNs/m}]$, and $\KPTO = -5~[\unit{kN/m}]$ and no power saturation limits. The WECs and their associated constraints are drawn in scale.
    Depiction of a single WEC with $2$x scaling is presented at the top of each plot for improved visualization.}
    \label{fig:LP_opt_none}
\end{figure*}

Overall, the WEC farm located on the Alaska Coast (East Coast) offers the largest (smallest) power and the largest (smallest) power per unit volume.
The optimized layout for each site location is shown in Fig.~\ref{fig:LP_opt_none}, where both WEC and their associated constraints are drawn in scale.
From these figures, it is clear that the optimized layout is different for each location.
This observation highlights the coupling between layout and site selection.
Furthermore, almost all of the optimized layouts resemble columns of WECs with a different number of clusters.
The horizontal separating distance among these columns is also different, with the smallest distance associated with the optimized layout on the East Coast, and the largest associated with the West Coast.  

\begin{table}[t]
\centering
\caption{Case Study \rom{3}: Concurrent geometric plant and layout optimization study for a $5$-WEC farm at different site locations using the hybrid optimization approach {(SM with \texttt{ga} + MS with \texttt{fmincon})} using $\BPTO = 500~[\unit{kNs/m}]$, and $\KPTO = -5~[\unit{kN/m}]$ and no power saturation limits.}
\label{Tab:LP_opt_Plim_none}

\begin{threeparttable}

\makebox[\linewidth]{%
\begin{tabular}{r r r r r r}
\hline  \hline
\multirow{2}{*}{\rotatebox[origin=c]{0}{\textbf{Site}}} & \multicolumn{2}{c}{\textrm{\textbf{Plant}}} &  \multirow{2}{*}{\textrm{\textbf{L}\tnote{b}}}& \multirow{2}{*}{\textrm{{P}\tnote{c}}} & \multirow{2}{*}{$q_\text{factor}$} \\ \cline{2-3}
& $\Rwec$\tnote{a} & $\RDwec$ & & &  \\
\hline 
\multirow{1}{*}{\rotatebox[origin=c]{0}{\textrm{NAWC24}}} & $3.03$ & $6.07$ & {\multirow{4}{*}{\rotatebox[origin=c]{90}{\textrm{Fig.~\ref{fig:LP_opt_none}}}}}& $167.19$ & $1.0112$ \\
\multirow{1}{*}{\rotatebox[origin=c]{0}{\textrm{NAEC8}}} & $3.59$ & $7.18$ & & $15.45$ & $1.0295$  \\
\multirow{1}{*}{\rotatebox[origin=c]{0}{\textrm{PI14}}} & $3.36$ & $6.71$ & & $35.00$ & $1.0287$  \\
\multirow{1}{*}{\rotatebox[origin=c]{0}{\textrm{N46229}}}  & $2.91$ & $5.83$ & & $71.33$ & $1.0085$ \\
\hline \hline
\end{tabular}%
}

\begin{tablenotes} [para,flushleft] \footnotesize
\item [a] Optimized in [\unit{m}]
\item [b] Optimized layout
\item [c] Power calculated over farm's lifetime using MS in [\unit{MW}]
\end{tablenotes}

\end{threeparttable}
\end{table}

\begin{figure}[t]
    \captionsetup[subfigure]{justification=centering}
    \centering
    \begin{subfigure}{\columnwidth}
    \centering
    \includegraphics[scale=\xywecscale]{Legend_LP_opt.pdf}
    \label{subfig:LP_opt_Legend_2}
    \end{subfigure}
    \begin{subfigure}{0.5\columnwidth}
    \centering
    \includegraphics[scale=\xywecscale]{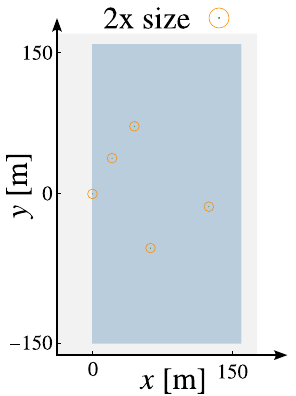}
    \caption{Alaska Coast (NAWC24).}
    \label{subfig:LP_opt_25_Alaska}
    \end{subfigure}%
    \begin{subfigure}{0.5\columnwidth}
    \centering
    \includegraphics[scale=\xywecscale]{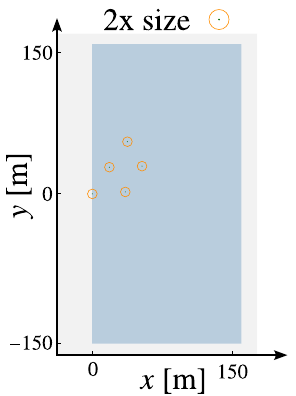}
    \caption{East Coast (NAEC8).}
    \label{subfig:LP_opt_25East}
    \end{subfigure}%
    \captionsetup[figure]{justification=centering}
    \caption{Case Study \rom{3}: Concurrent geometric plant and layout optimization study for a $5$-WEC farm using the hybrid optimization approach {(SM with \texttt{ga} + MS with \texttt{fmincon})} with $\BPTO = 3~[\unit{kNs/m}]$, and $\KPTO = -5~[\unit{kN/m}]$, $p_{\text{lim}} = 25~[\unit{kW}]$ saturation limit at the Alaska Coast and East Coast. The WECs and their associated constraints are drawn in scale. 
    Depiction of a single WEC with $2$x scaling is presented at the top of each plot for improved visualization.}
    \label{fig:LP_opt_25}
\end{figure}

To investigate the impact of power saturation limits on concurrent plant and control optimization, a similar study is carried out with $\BPTO = 3~[\unit{kNs/m}]$, and $\KPTO = -5~[\unit{kN/m}]$ and $p_{\text{lim}} = 25~[\unit{kW}]$ for the Alaska Coast (NAWC24) and East Coast (NAEC8).
The results from this investigation are presented in Table~\ref{Tab:LP_opt_Plim_25k} and Fig.~\ref{fig:LP_opt_25}.
According to Table~\ref{Tab:LP_opt_Plim_25k}, when a power saturation limit of $p_{\text{lim}}= 25~[\unit{kW}]$ is imposed, the WEC radius and draft remain at their lower bound for both locations, indicating that with stringent limitations on control effort, small devices are more suited for producing higher power per unit volume. 
As evidenced by Fig.~\ref{fig:LP_opt_25}, small device dimensions directly impact the optimized layout of the farm.
Compared to solutions presented in Fig.~\ref{fig:LP_opt_none}, the optimized layouts are more centrally located.
In the case of the East Coast, the WEC devices are closely packed in a single region better to leverage the constructive interaction effects among these small devices.  

\begin{table}[t]
\centering
\caption{Case Study \rom{3}: Concurrent geometric plant and layout optimization study for a $5$-WEC farm at different site locations using the hybrid optimization approach {(SM with \texttt{ga} + MS with \texttt{fmincon})} using $\BPTO = 3~[\unit{kNs/m}]$, and $\KPTO = -5~[\unit{kN/m}]$ and $p_{\text{lim}} = 25~[\unit{kW}]$ saturation limit.}
\label{Tab:LP_opt_Plim_25k}

\begin{threeparttable}

\makebox[\linewidth]{%
\begin{tabular}{r r r r r r}
\hline  \hline
\multirow{2}{*}{\rotatebox[origin=c]{0}{\textbf{Site}}} & \multicolumn{2}{c}{\textrm{\textbf{Plant}}} &  \multirow{2}{*}{\textrm{\textbf{L}\tnote{b}}}& \multirow{2}{*}{\textrm{{P}\tnote{c}}} & \multirow{2}{*}{$q_\text{factor}$} \\ \cline{2-3}
& $\Rwec$\tnote{a} & $\RDwec$ & & & \\
\hline 
\multirow{1}{*}{\rotatebox[origin=c]{0}{\textrm{NAWC24}}} & $0.50$ & $1.00$ & {\multirow{2}{*}{\rotatebox[origin=c]{0}{\textrm{Fig.~\ref{fig:LP_opt_25}}}}} & $1.64$ & $1.0001$ \\
\multirow{1}{*}{\rotatebox[origin=c]{0}{\textrm{NAEC8}}} & $0.50$ & $1.00$ & & $0.44$ & $1.0026$  \\
\hline \hline
\end{tabular}%
}%

\begin{tablenotes} [para,flushleft] \footnotesize
\item [a] Optimized in [\unit{m}]
\item [b] Optimized layout
\item [c] Power calculated over farm's lifetime using MS in [\unit{MW}] 
\end{tablenotes}
     
\end{threeparttable}
\end{table}

This investigation creates an opportunity to compare the proposed framework to similar works in the literature.
As an example, Lyu et al. conducted a concurrent optimization of WEC dimension and layout for arrays of $3$, $5$, and $7$ WECs using GA~\citep{lyu2019optimization}.
However, due to high computational costs, only a single study with $3$ WECs was performed in irregular waves, which was truncated after $7$ GA generations.
Evidenced by the results reported in this section, using the proposed framework, a similar study for a farm consisting of $5$ WEC devices is successfully solved with GA in the presence of probabilistic irregular waves in less than $2~[\unit{h}]$.

Consequently, the proposed approach improves the computational efficiency of WEC design optimization studies tremendously, ultimately enabling investigations with additional complexity, including all aforementioned design decision areas and more individual WECs, which are lacking in the WEC literature to the best of the authors' knowledge.
These investigations are carried out next.

\subsubsection{Case Study \rom{4}: Concurrent Geometric Plant, Farm-level Control, and Layout Optimization}
\label{subsubsec:CaseStudy4}

\begin{figure*}[t]
    \captionsetup[subfigure]{justification=centering}
    \centering
    \begin{subfigure}{\textwidth}
    \centering
    \includegraphics[scale=\xywecscale]{Legend_LP_opt.pdf}
    \label{subfig:PCL_farm_Legend_1}
    \end{subfigure}
    \begin{subfigure}{0.25\textwidth}
    \centering
    \includegraphics[scale=\xywecscale]{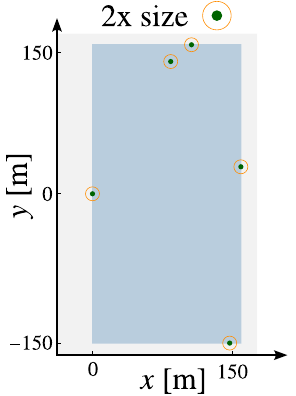}
    \caption{Alaska Coast (NAWC24).}
    \label{subfig:PCL_farm_none_Alaska}
    \end{subfigure}%
    \begin{subfigure}{0.25\textwidth}
    \centering
    \includegraphics[scale=\xywecscale]{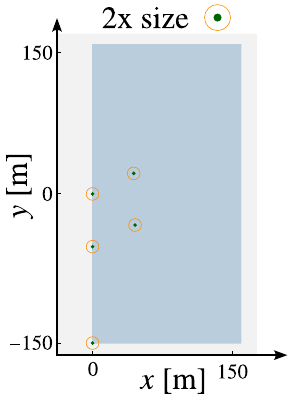}
    \caption{East Coast (NAEC8).}
    \label{subfig:PCL_farm_none_East}
    \end{subfigure}%
    \begin{subfigure}{0.25\textwidth}
    \centering
    \includegraphics[scale=\xywecscale]{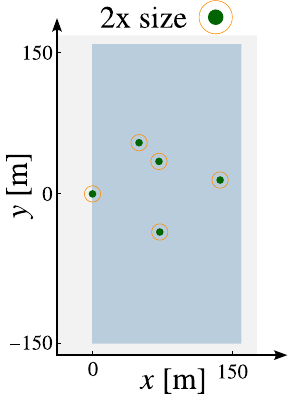}
    \caption{Pacific Islands (PI14).}
    \label{subfig:PCL_farm_none_Pacific}
    \end{subfigure}%
    \begin{subfigure}{0.25\textwidth}
    \centering
    \includegraphics[scale=\xywecscale]{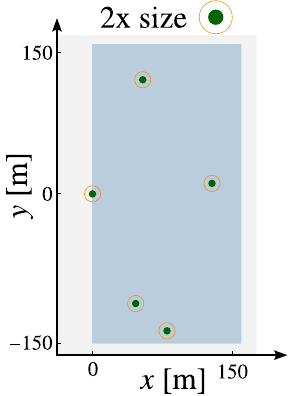}
    \caption{West Coast (N46229).}
    \label{subfig:PCL_farm_none_West}
    \end{subfigure}%
    \captionsetup[figure]{justification=centering}
    \caption{Case Study \rom{4}: Concurrent geometric plant, farm-level control, and layout optimization study for a $5$-WEC farm at different site locations using the hybrid optimization approach {(SM with \texttt{ga} + MS with \texttt{fmincon})} with no power saturation limits.  The WECs and their associated constraints are drawn in scale.
    Depiction of a single WEC with $2$x scaling is presented at the top of each plot for improved visualization.}
    \label{fig:PCL_farm_none}
\end{figure*}

Moving forward with increasing complexity and problem size, this case study implements concurrent plant, farm-level control, and layout optimization.
In this description, farm-level control refers to the determination of the optimal PTO control parameters across the farm by the optimizer, such that all WEC devices within the farm share the same PTO design ($\bm{u} \in \mathbb{R}^{{2}}$).
The inclusion of farm-level control optimization, therefore, adds $2$ additional variables to the set of optimization variables, resulting in a total of $12$ optimization variables for a $5$-WEC farm.   
Similar to the previous case, this study also examines the impact of power saturation limits on the optimal integrated solution.

The results from the investigation without any power saturation limits are presented in Table~\ref{Tab:PCL_farm_None} and Fig.~\ref{fig:PCL_farm_none}.
In all of the cases studied so far, we observed that the WEC radius is generally larger for the East Coast, which is a low wave energy resource region. 
The results presented in Table~\ref{Tab:PCL_farm_None}, however, indicate that the WEC radius on the East Coast is the smallest (and the one associated with the Pacific Islands is the largest).   
This observation is directly related to the role and impact of controller design on plants.
Specifically, when no power saturation limits are imposed, the optimal resonant controller amplifies the device's motion and power generation.
This outcome is evident from the fact that the power generated by the farm in this study is much larger than Case Study \rom{3}, when only plant and layout variables were optimized. 
Since controller optimality has a significant impact on power generation, the previous trends are no longer observed as the coupling between plant and control domains is now signified.   
This observation highlights the critical impact of domain coupling and integrated studies in early-stage design of such complex systems.

In addition, these results indicate some variations in the WEC draft dimension, a behavior that was not observed in the previous case studies.
Specifically, the WEC draft is $0.53$, $0.50$, $0.6$, and $0.59~[\unit{m}]$ for Alaska Coast (NAWC24), East Coast (NAEC8), Pacific Islands (PI14), and West Coast (N46229), respectively.  
According to Table~\ref{Tab:PCL_farm_None}, the optimal control parameters also change for each location.
While $\BPTO$ remains relatively small, $\KPTO$ is large in magnitude, relatively close to its lower bound, and negative, enabling a reactive phase control \cite{antonio2010wave} strategy that can extend the range of resonance conditions \citep{todalshaug2016tank, peretta2015effect}.

\begin{table}[t]
\centering
\caption{Case Study \rom{4}: Concurrent geometric plant, farm-level control, and layout optimization study for a $5$-WEC farm at different site locations using the hybrid optimization approach {(SM with \texttt{ga} + MS with \texttt{fmincon})} with no power saturation limits.}
\label{Tab:PCL_farm_None}

\begin{threeparttable}

\makebox[\linewidth]{%
\small
\begin{tabular}{r r r r r r r }
\hline  \hline
\multirow{2}{*}{\textrm{\textbf{Site}}} & \multicolumn{2}{c}{\textrm{\textbf{Plant}}} &  \multicolumn{2}{c}{\textrm{\textbf{Control}\tnote{b}}}& \multirow{2}{*}{\textrm{\textbf{L}\tnote{c}}} & \multirow{2}{*}{\textrm{{P}}\tnote{d}}    \\ \cline{2-3} \cline{4-5}
& $\Rwec \tnote{a}$ & $\RDwec$ & $\BPTO$ & $\KPTO$  & & \\
  \hline 
$\text{NAWC24}$ & $2.79$ & $5.29$ & $0.92$ & $-226.01$ &  \multirow{4}{*}{\rotatebox[origin=c]{90}{\textrm{Fig.~\ref{fig:PCL_farm_none}}}} & $9.7\times 10^{7}$   \\
$\text{NAEC8}$ & $2.11$ & $4.22$ & $1.40$ & $-116.00$ &  & {$4.2 \times10^{1}$} \\ 
$\text{PI14}$ & $4.06$ & $6.80$ & $13.84$& $-419.47$ &  & $2.9\times10^6$   \\
$\text{N46229}$ & $3.99$ & $6.79$ & $3.62$& $-434.66$ &  & $5.2 \times 10^{7}$   \\
\hline \hline
\end{tabular}%
}

\begin{tablenotes} [para,flushleft]  \footnotesize
\item [a] Calculated in [\unit{m}]
\item [b] Calculated in [\unit{kNs/m}] and [\unit{kN/m}]
\item [c] Optimized layout
\item [d] Calculated power in [\unit{TW}] over farm lifetime using MS  
\end{tablenotes}

\end{threeparttable}
\end{table}

The optimized layouts for this study are shown in Fig.~\ref{fig:PCL_farm_none}.
As opposed to the optimized layout configurations presented in Fig.~\ref{fig:LP_opt_none}, the optimized layouts for this study are less regular and, at least for the East Coast (NAEC8) and Pacific Islands (PI14), more centrally clustered.  
The variations witnessed in these results, compared to the previous studies, point to the coupling that exists between plant and layout optimization. 

Note that the results presented here do not include an estimation of the $q_\text{factor}$. 
This decision is because when control optimization is involved, the optimal controller is often a resonator, resulting in large amplitudes of the device motion.
The natural frequency of WECs is then defined as:
\begin{equation}
    \omega_n = \sqrt{\frac{\bm{k}_{\text{pto}} + G}{\mathbf{M} + \mathbf{A}(\omega,\mathbf{w})}}
\end{equation}
\noindent
where the added mass $\AddedMass$ entails contributions from surrounding devices.
This equation implies that the optimal control parameters are optimized for the farm but are off-resonance for individual WEC devices.
This observation points to some of the limitations of the $q_\text{factor}$ for integrated studies such as CCD because the resulting value compares the farm performance with a resonance controller in the numerator to a device performance with an off-resonance controller in the denominator.

\begin{figure}[t]
    \captionsetup[subfigure]{justification=centering}
    \centering
    \begin{subfigure}{\columnwidth}
    \centering
    \includegraphics[scale=\xywecscale]{Legend_LP_opt.pdf}
    \label{subfig:PCL_farm_Legend_2}
    \end{subfigure}
    \begin{subfigure}{0.5\columnwidth}
    \centering
    \includegraphics[scale=\xywecscale]{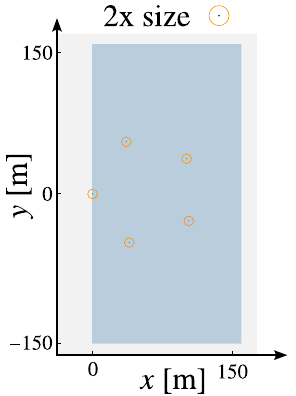}
    \caption{Alaska Coast (NAWC24).}
    \label{subfig:PCL_Farm_75_Alaska}
    \end{subfigure}%
    \begin{subfigure}{0.5\columnwidth}
    \centering
    \includegraphics[scale=\xywecscale]{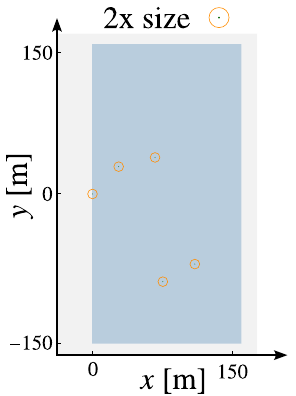}
    \caption{East Coast (NAEC8).}
    \label{subfig:PCL_Farm_75_East}
    \end{subfigure}%
    \captionsetup[figure]{justification=centering}
    \caption{Case Study \rom{4}: Concurrent geometric plant, farm-level control, and layout optimization study for a $5$-WEC farm using the hybrid optimization approach {(SM with \texttt{GA} + MS with \texttt{fmincon})} with power saturation limit of $p_{\text{lim}} = 75~[\unit{kW}]$ at Alaska Coast and East Coast. The WECs and their associated constraints are drawn in scale.
    Depiction of a single WEC with $2$x scaling is presented at the top of each plot for improved visualization.}
    \label{fig:PCL_Farm_75}
\end{figure}

The impact of power saturation limits is considered next.
Assuming the prescribed power saturation limit of $p_{\text{lim}} = 75~[\unit{kW}]$, the results are for the Alaska Coast (NAEC24) and East Coast (NAEC8) are presented in Table~\ref{Tab:PCL_farm_75k} and Fig.~\ref{fig:PCL_Farm_75}. 
Once again, in the presence of power saturation limits the geometry of the WEC device tends towards smaller devices with both the radius and draft dimensions at their lower bound.
As expected, the power captured by the farm is extremely reduced compared to the case with no power saturation limits, with the site on the Alaska Coast (NAEC24) still producing higher levels of power.
The optimized layout for the Alaska Coast highlights a relatively symmetrical solution with WECs clustered at the center of the farm (vertically).

\begin{table}[t]
\centering
\caption{Case Study \rom{4}: Concurrent geometric plant, farm-level control, and layout optimization study for a $5$-WEC farm using the hybrid optimization approach {(SM with \texttt{GA} + MS with \texttt{fmincon})} with power saturation limit of $p_{\text{lim}} = 75~[\unit{kW}]$ at Alaska Coast and East Coast.}
\label{Tab:PCL_farm_75k}

\begin{threeparttable}

\makebox[\linewidth]{%
\begin{tabular}{r r r r r r r }
\hline  \hline
\multirow{2}{*}{\textrm{\textbf{Site}}} & \multicolumn{2}{c}{\textrm{\textbf{Plant}}} &  \multicolumn{2}{c}{\textrm{\textbf{Control}\tnote{b}}}& \multirow{2}{*}{\textrm{\textbf{L}\tnote{c}}} & \multirow{2}{*}{\textrm{{P}}\tnote{d}}    \\ \cline{2-3} \cline{4-5}
& $\Rwec \tnote{a}$ & $\RDwec$ & $\BPTO$ & $\KPTO$  & & \\
  \hline 
$\text{NAWC24}$ & $0.50$ & $1.00$ & $0.15$ & $-7.48$ &  \multirow{2}{*}{\rotatebox[origin=c]{0}{\textrm{Fig.\ref{fig:PCL_Farm_75}}}} & $6.30$   \\
$\text{NAEC8}$ & $0.50$ & $1.00$ & $0.18$ & $-6.98$ &  & $1.69$  \\
\hline \hline
\end{tabular}%
}

\begin{tablenotes} [para,flushleft]  \footnotesize
\item [a] Calculated in [\unit{m}]
\item [b] Calculated in [\unit{kNs/m}] and [\unit{kN/m}]
\item [c] Optimized layout
\item [d] Power calculated in [\unit{MW}] over the lifetime of the farm using MS  
\end{tablenotes}

\end{threeparttable}
\end{table}

\subsubsection{Case Study \rom{5}: Concurrent Plant, Device-level Control, and Layout Optimization}
\label{subsubsec:CaseStudy5}

As opposed to the previous study where one set of PTO control parameters were optimized and used for the entire farm, this study investigates a concurrent optimization problem with device-level control optimization. 
Therefore, the PTO parameters for each WEC within the farm are selected individually by the optimizer, such that $\bm{u} \in \mathbb{R}^{{2}\Nwec}$.
Theoretically, this approach is expected to result in an increase in the power per unit volume of the farm.
However, the number of optimization variables increases by $2(n_{wec}-1)$, compared to the previous study with farm-level control, resulting in a total of $20$ decision variables.
This study also examines the impact of power saturation limits on the optimized solution.

First, the results from this investigation with no power saturation limits are presented in Table~\ref{Tab:PCL_device_None}.
According to Table~\ref{Tab:PCL_device_None} the optimized power of the farm for the Alaska Coast (NAWC24) and the Pacific Islands (PI14) exceeds the power obtained in Case Study \rom{4}, where a farm-level control optimization was utilized.
However, the farm power for the East Coast (NAEC8) and the West Coast (N46229) in the concurrent optimization with device-level control is lower than that of the farm-level control in Case Study \rom{4}.
This trend also holds for the objective function.
This observation, which might be related to the increase in the size of the optimization problem (and thus challenges in finding superior solutions), requires further investigation. 
From these results, it is also clear that the optimal plant values have changed compared to Case Study \rom{4}.
While there are no regular trends observed, these variations highlight the potential benefits of using device-level control optimization to improve farm performance. 
Similar to the previous case, the draft dimension, which remained at $0.5~[\unit{m}]$ in the first three case studies, undergoes variations with $0.56~[\unit{m}]$ for Alaska Coast (NAWC24), $0.61~[\unit{m}]$ for the East Coast (NAEC8), $0.54~[\unit{m}]$ for the Pacific Islands (PI14), and $0.60~[\unit{m}]$ for the West Coast (N46229).

\begin{table}[t]
\centering
\caption{Case Study \rom{5}: Concurrent geometric plant, device-level control, and layout optimization study for a $5$-WEC farm at various site locations using the hybrid optimization approach {(SM with \texttt{ga} + MS with \texttt{fmincon})} with no power saturation limits.}
\label{Tab:PCL_device_None}

\begin{threeparttable}[b]

\makebox[\linewidth]{%
\begin{tabular}{r r r r r r}
\hline  \hline
\multirow{2}{*}{\textrm{\textbf{Site}}} & \multicolumn{2}{c}{\textrm{\textbf{Plant}}} &  \multirow{2}{*}{\textrm{\textbf{Control}}}& \multirow{2}{*}{\textrm{\textbf{L}\tnote{c}}} & \multirow{2}{*}{\textrm{{P}}\tnote{d}}    \\ \cline{2-3} 
& $\Rwec \tnote{a}$ & $\RDwec$ &  & & \\
  \hline 
$\text{NAWC24}$ & $3.92$ & $6.97$ & \multirow{4}{*}{\rotatebox[origin=c]{90}{\textrm{Fig.~\ref{subfig:device_level_control_None}}}} &  \multirow{4}{*}{\rotatebox[origin=c]{90}{\textrm{Fig.~\ref{fig:PCL_device_none}}}} & $420.4\times 10^7$   \\
$\text{NAEC8}$ & $4.11$ & $6.76$ &  &  & $3.2\times 10^{-1}$   \\ 
$\text{PI14}$ & $3.79$ & $7.08$ &  &  & $3.3\times 10^6$   \\
$\text{N46229}$ & $3.25$ & $5.45$ &  &  & $1.9\times 10^4$   \\
\hline \hline
\end{tabular}%
}

\begin{tablenotes} [para,flushleft]  \footnotesize
\item [a] Calculated in [\unit{m}]
\item [c] Optimized layout \item
[d] Calculated power in [\unit{TW}] over farm lifetime using MS  
\end{tablenotes}

\end{threeparttable}
\end{table}

\begin{figure}
    \captionsetup[subfigure]{justification=centering}
    \centering
    \begin{subfigure}{0.5\columnwidth}
    \centering
    \includegraphics[scale=\twofigscale]{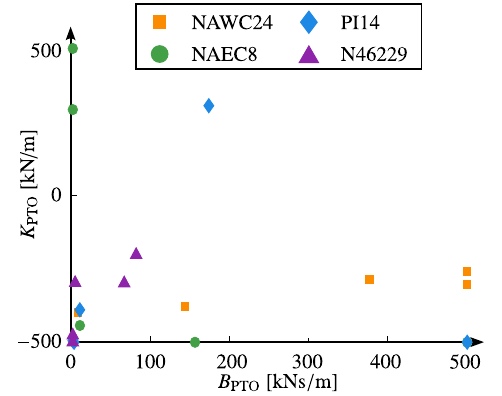}
    \caption{No power limits.}
    \label{subfig:device_level_control_None}
    \end{subfigure}%
    \begin{subfigure}{0.5\columnwidth}
    \centering
    \includegraphics[scale=\twofigscale]{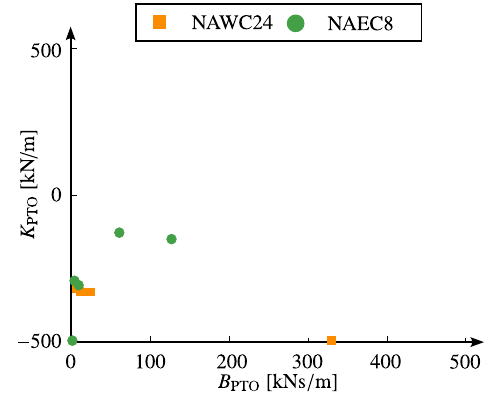}
    \caption{Power limit $p_{\text{lim}} = 100~[\unit{MW}]$.}
    \label{subfig:device_level_control_100MW}
    \end{subfigure}
    \captionsetup[figure]{justification=centering}
    \caption{Optimized device-level PTO control parameters for concurrent plant, device-level control, and layout optimization of Case Study \rom{5}.}
    \label{fig:device_level_control}
\end{figure}

\begin{figure*}[t]
    \captionsetup[subfigure]{justification=centering}
    \centering
    \begin{subfigure}{\columnwidth}
    \centering
    \includegraphics[scale=\xywecscale]{Legend_LP_opt.pdf}
    \label{subfig:PCL_device_Legend_1}
    \end{subfigure}
    \begin{subfigure}{0.25\textwidth}
    \centering
    \includegraphics[scale=\xywecscale]{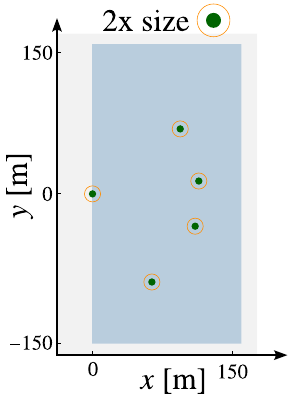}
    \caption{Alaska Coast (NAWC24).}
    \label{subfig:PCL_device_none_Alaska}
    \end{subfigure}%
    \begin{subfigure}{0.25\textwidth}
    \centering
    \includegraphics[scale=\xywecscale]{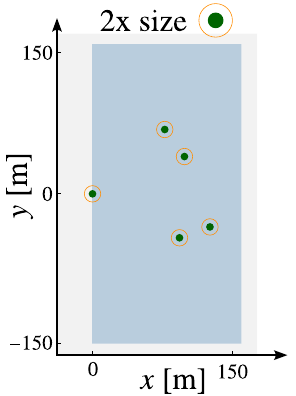}
    \caption{East Coast (NAEC8).}
    \label{subfig:PCL_device_none_East}
    \end{subfigure}%
    \begin{subfigure}{0.25\textwidth}
    \centering
    \includegraphics[scale=\xywecscale]{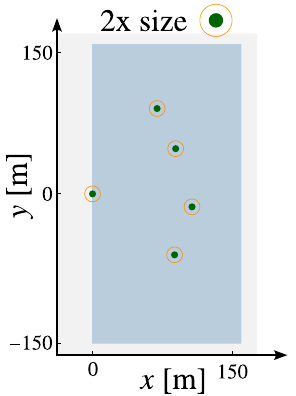}
    \caption{Pacific Islands (PI14).}
    \label{subfig:PCL_device_none_Pacific}
    \end{subfigure}%
    \begin{subfigure}{0.25\textwidth}
    \centering
    \includegraphics[scale=\xywecscale]{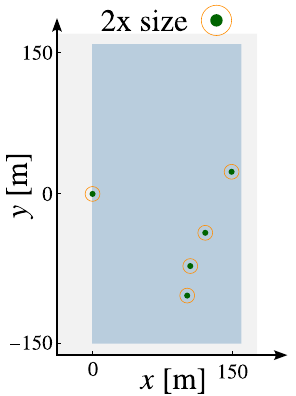}
    \caption{West Coast (N46229).}
    \label{subfig:PCL_device_none_West}
    \end{subfigure}%
    \captionsetup[figure]{justification=centering}
    \caption{Case Study \rom{5}: Concurrent geometric plant, device-level control, and layout optimization study for a $5$-WEC farm at various site locations using the hybrid optimization approach {(SM with \texttt{ga} + MS with \texttt{fmincon})} with no power saturation limits. The WECs and their associated constraints are drawn in scale.
    Depiction of a single WEC with $2$x scaling is presented at the top of each plot for improved visualization.}
    \label{fig:PCL_device_none}
\end{figure*}

The device-level optimized control variables for this study are shown in Fig.~\ref{subfig:device_level_control_None}.
According to this figure, at least for two of the WEC devices on the East Coast (NAEC8), the optimized PTO stiffness is positive.
This result might limit the range of resonance in reactive phase control and seems to be a culprit in lower power generation for the East Coast (NAEC8) in Study \rom{5} compared to \rom{4}. 
The optimized layout of the farm is shown in Fig.~\ref{fig:PCL_device_none}.
Clearly, the optimized layout of the farm has changed compared to the case Study \rom{4}.
Specifically, the optimized layout in the Alaska Coast (NAWC24) and the Pacific Islands (PI14) are more regular than those associated with the East Coast (NAEC8) and the West Coast (N46229). 

\begin{figure}[t]
    \captionsetup[subfigure]{justification=centering}
    \centering
    \begin{subfigure}{\columnwidth}
    \centering
    \includegraphics[scale=\xywecscale]{Legend_LP_opt.pdf}
    \label{subfig:PCL_device_Legend_2}
    \end{subfigure}
    \begin{subfigure}{0.5\columnwidth}
    \centering
    \includegraphics[scale=\xywecscale]{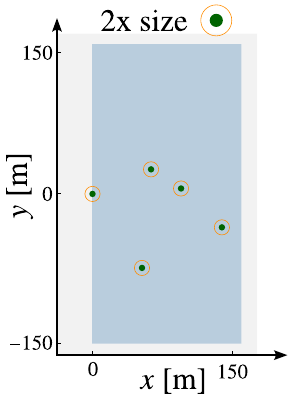}
    \caption{Alaska Coast (NAWC24).}
    \label{subfig:PCL_device_75_Alaska}
    \end{subfigure}%
    \begin{subfigure}{0.5\columnwidth}
    \centering
    \includegraphics[scale=\xywecscale]{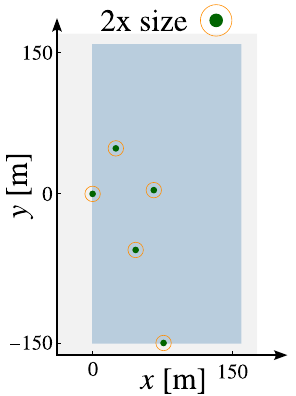}
    \caption{East Coast (NAEC8).}
    \label{subfig:PCL_device_75_East}
    \end{subfigure}%
    \captionsetup[figure]{justification=centering}
    \caption{Case Study \rom{5}: Optimized layout associated with the concurrent geometric plant, device-level control, and layout optimization study for a $5$-WEC farm at Alaska Coast and East Coast locations using the hybrid optimization approach {(SM with \texttt{ga} + MS with \texttt{fmincon})} with power saturation limit of $p_{\text{lim}} = 100~[\unit{MW}]$.
    The WECs and their associated constraints are drawn in scale.
    Depiction of a single WEC with $2$x scaling is presented at the top of each plot for improved visualization.}
    \label{fig:PCL_Device_100MW}
\end{figure}

The impact of power saturation limits is studied next with the assumption of $p_{\text{lim}} = 100~[\unit{MW}]$.
According to Table~\ref{Tab:PCL_device_100M}, the WEC device located on the site at the East Coast (NAEC8) location has a larger radius and a slightly smaller draft value. 
The optimized control parameters and the layout from this study are also shown in Figs.~\ref{subfig:device_level_control_100MW} and \ref{fig:PCL_Device_100MW}, respectively. 
Similar to the previous cases, the optimized layout has changed compared to the case without power saturation limits.  
The PTO stiffness parameters for the East Coast, as witnessed by Fig.~\ref{subfig:device_level_control_100MW}, are now all negative, indicating that the optimized controller takes advantage of the potential benefits of the reactive phase control. 
The optimized layouts shown in Fig.~\ref{fig:PCL_Device_100MW} seem more centrally clustered compared to results with no power saturation limits.

\begin{table}[t]
\centering
\caption{Case Study \rom{5}: Concurrent geometric plant, device-level control, and layout optimization study for a $5$-WEC farm at Alaska Coast and East Coast locations using the hybrid optimization approach {(SM with \texttt{ga} + MS with \texttt{fmincon})} with power saturation limit of $p_{\text{lim}} = 100~[\unit{MW}]$.}
\label{Tab:PCL_device_100M}

\begin{threeparttable}

\makebox[\linewidth]{%
\begin{tabular}{r r r r r r}
 \hline  \hline
\multirow{2}{*}{\textrm{\textbf{Site}}} & \multicolumn{2}{c}{\textrm{\textbf{Plant}}} &  \multirow{2}{*}{\textrm{\textbf{Control}\tnote{b}}}& \multirow{2}{*}{\textrm{\textbf{L}\tnote{c}}} & \multirow{2}{*}{\textrm{{P}}\tnote{d}}    \\ \cline{2-3} 
 & $\Rwec \tnote{a}$ & $\RDwec$ & & & \\
      \hline 
 $\text{NAWC24}$ & $3.44$ & $6.70$ & \multirow{2}{*}{\rotatebox[origin=c]{0}{\textrm{Fig.~\ref{subfig:device_level_control_100MW}}}} &  \multirow{2}{*}{\rotatebox[origin=c]{0}{\textrm{Fig.~\ref{fig:PCL_Device_100MW}}}} & $3.44$   \\
 $\text{NAEC8}$ & $3.56$ & $7.11$ &  &  & $0.97$  \\
\hline \hline
\end{tabular}%
}%

\begin{tablenotes} [para,flushleft]  \footnotesize
\item [a] Calculated in [\unit{m}]
\item [b] Calculated in [\unit{kNs/m}] and [\unit{kN/m}]
\item [c] Optimized layout
\item [d] Power calculated in [\unit{GW}] over the lifetime of the farm using MS  
\end{tablenotes}

\end{threeparttable}
\end{table}

\subsubsection{Case Study \rom{6}: Concurrent Plant, Farm-level Control, and Layout Optimization for a $25$-WEC Farm}
\label{subsubsec:CaseStudy6}

\begin{table}[t]
\centering
\caption{Optimized solution for Study~\rom{6}: Concurrent plant, farm-level control, and layout optimization with power saturation limit of $p_{\text{lim}} = 100~[\unit{MW}]$ at Alaska Coast and East Coast using a hybrid optimization approach {(SM with \texttt{ga} + MS with \texttt{fmincon})} for a $25$-WEC farm.}
\label{Tab:PCL_farm_100MW_25WEC}

\begin{threeparttable}

\makebox[\linewidth]{%
\small
\begin{tabular}{r r r r r r r }
\hline  \hline
\multirow{2}{*}{\textrm{\textbf{Site}}} & \multicolumn{2}{c}{\textrm{\textbf{Plant}}} &  \multicolumn{2}{c}{\textrm{\textbf{Control}\tnote{b}}}& \multirow{2}{*}{\textrm{\textbf{L}\tnote{c}}} & \multirow{2}{*}{\textrm{{P}}\tnote{d}}    \\ \cline{2-3} \cline{4-5}
& $\Rwec \tnote{a}$ & $\RDwec$ & $\BPTO$ & $\KPTO$ & & \\
  \hline 
$\text{NAWC24}$ & $2.58$ & $5.10$ & $3.46$ & $-188.9$ &  \multirow{2}{*}{\rotatebox[origin=c]{0}{\textrm{Fig.\ref{fig:PCL_Device_100MW_25}}}} & $11.33$   \\
$\text{NAEC8}$ & $4.76$ & $9.37$ & $133.91$ & $-475.3$ &  & $0.76$  \\
\hline \hline
\end{tabular}%
}

\begin{tablenotes} [para,flushleft]  \footnotesize
\item [a] Calculated in [\unit{m}]
\item [b] Calculated in [\unit{kNs/m}] and [\unit{kN/m}] \item [c] Optimized layout
\item [d] Power calculated in [\unit{GW}] over the lifetime of the farm using MS  
\end{tablenotes}

\end{threeparttable}
\end{table}

\begin{figure}[t]
    \captionsetup[subfigure]{justification=centering}
    \centering
    \begin{subfigure}{\columnwidth}
    \centering
    \includegraphics[scale=\xywecscale]{Legend_LP_opt.pdf}
    \label{subfig:PCL_Farm_Legend_25}
    \end{subfigure}
    \begin{subfigure}{0.5\columnwidth}
    \centering
    \includegraphics[scale=\xywecscale]{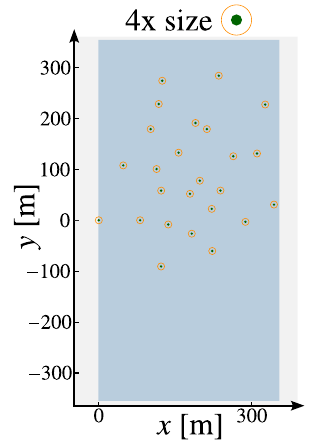}
    \caption{Alaska Coast (NAWC24).}
    \label{subfig:PCL_Farm_100MW_Alaska_25WEC}
    \end{subfigure}%
    \begin{subfigure}{0.5\columnwidth}
    \centering
    \includegraphics[scale=\xywecscale]{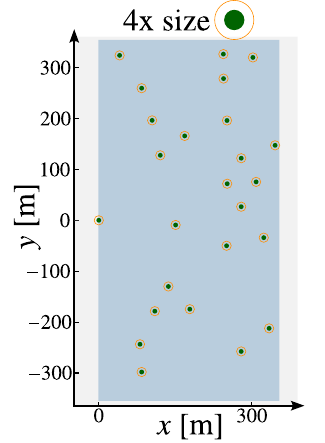}
    \caption{East Coast (NAEC8).}
    \label{subfig:PCL_Farm_100MW_East_25WEC}
    \end{subfigure}%
    \captionsetup[figure]{justification=centering}
    \caption{Optimized layout from Case Study \rom{6}: Concurrent plant, farm-level control, and layout optimization with power saturation limit of $p_{\text{lim}} = 100~[\unit{MW}]$ for a $25$-WEC farm on the Alaska Coast and East Coast using a hybrid optimization approach {(SM with \texttt{ga} + MS with \texttt{fmincon})}. The WECs and their associated constraints are drawn in scale.
    Depiction of a single WEC with $4$x scaling is presented at the top of each plot for improved visualization.}
    \label{fig:PCL_Device_100MW_25}
\end{figure}

With the goal of answering some questions about the scalability of the proposed approach, this study focuses on investigating the concurrent plant, farm-level control, and layout optimization for a WEC farm with $25$ WEC devices.
Using a power saturation limit of $p_{\text{lim}} = 100~[\unit{MW}]$, the problem is solved at the Alaska Coast (NAWC24) and East Coast (NAEC8).
Similar to the established trend, we first solve the concurrent CCD problem using SMs, and then use the optimized solution with a gradient-based optimizer and MS to find a more accurate solution.
The results from this study are presented in Table~\ref{Tab:PCL_farm_100MW_25WEC}, and Fig.~\ref{fig:PCL_Device_100MW_25}, on a high-performance computer with $40$ cores on $2~\times~$Intel Xeon Gold $6148$ CPUs and $192~[\unit{GB}]$ memory.   
The SM solutions were obtained within $17~[\unit{h}]$, and the hybrid optimization solution took approximately $76~[\unit{h}]$ to converge, resulting in an overall computational time of approximately $93~[\unit{h}]$ or $3.87~[\unit{days}]$.   
Comparing this solution with the $639~[\unit{h}]$ required to solve the layout optimization problem of a $5$-WEC farm using MS, it is evident that this approach is promising in enabling more complex integrated studies of WEC farms with high number of WEC devices.

\section{Conclusion}\label{sec:conclusion}

Integrated design approaches based on control co-design (CCD) constitute the concurrent consideration of geometric plant, control, and layout optimization in order to improve the performance and economic competitiveness of WEC farms.
The bottleneck in such studies is the estimation of hydrodynamic coefficients.
Therefore, in this article, we used the principles of many-body expansion (MBE) to develop data-driven artificial neural networks (ANNs) based on an active learning strategy known as query by committee to estimate the hydrodynamic coefficients.

The effectiveness of the surrogate models was validated through a variety of simulations and optimization studies, and consolidated by employing a hybrid optimization strategy. 
The heaving WEC model was developed in the frequency domain, with WEC radius and slenderness ratio as plant optimization variables, power take-off control damping and stiffness as control optimization variables, and center coordinates of WEC devices as layout optimization variables.
Next, a series of optimization problems with increasing levels of integration and complexity was introduced and solved at various site locations, culminating in a concurrent geometric plant, control, and layout optimization of a $25$-WEC farm.

The proposed approach enables a more efficient implementation of single-domain investigations.
More importantly, it enables investigations that to the best of the authors' knowledge are currently lacking from the literature due to their high computational cost. 
The study and results also highlight the role of domain coupling in obtaining superior WEC farm solutions.

While the proposed approach and surrogate models offer promising directions for more integrated investigations, it is important to note that the results presented in this article are based on some specific assumptions.
For example, WEC devices were modeled in frequency domain, in the absence of any direct and explicit constraint on device motion.
Mooring costs, electrical interconnections, and power smoothing properties were excluded. 
In addition, we terminated the MBE series at the second term.
However, utilizing a higher MBE order enables a more accurate characterization of the hydrodynamic interaction effect, potentially making the approach more accurate for larger WEC farms.

An important future work is the quantification of the epistemic uncertainties resulting from surrogate modeling.
Utilizing a more practical controller with appropriate constraints (such as device motion, PTO force, etc.) and assessing the proposed approach with time-domain models might provide additional insights to pave the way for improving the integrated design of WEC farms.

\section*{Acknowledgment}
\label{sec:Acknowledgment}
The authors gratefully acknowledge the financial support from National Science Foundation Engineering Design and Systems Engineering Program, USA under grant number CMMI-2034040.










\renewcommand{\refname}{REFERENCES}
\bibliographystyle{Config/elsarticle-num-names}
\bibliography{references}





\end{document}

%% file: Front.tex
\begin{frontmatter}



\title{Concurrent Geometry, Control, and Layout Optimization of Wave Energy Converter Farms in Probabilistic Irregular Waves using Surrogate Modeling} 


\author[first]{Saeed Azad\corref{cor1}} 
\ead{saeed.azad@colostate.edu}

\author[first]{Daniel R. Herber}
\ead{daniel.herber@colostate.edu}

\author[second]{Suraj Khanal}
\ead{Suraj.Khanal@colostate.edu}

\author[second]{Gaofeng Jia}
\ead{gjia@colostate.edu}

\affiliation[first]{organization={Systems Engineering Department, Colorado State University},
            city={Fort Collins},
            state={CO},
            country={USA}}

\affiliation[second]{organization={Civil \& Environmental Engineering Department, Colorado State University}, 
            city={Fort Collins},
            state={CO},
            country={USA}}
            
\cortext[cor1]{Corresponding author}

\begin{abstract}
A promising direction towards improving the performance of wave energy converter (WEC) farms is to leverage a system-level integrated approach known as control co-design (CCD).
A WEC farm CCD problem may entail decision variables associated with the geometric attributes, control parameters, and layout of the farm.
However, solving the resulting optimization problem, which requires the estimation of hydrodynamic coefficients through numerical methods such as multiple scattering (MS), is computationally prohibitive. 
To mitigate this computational bottleneck, in this article, we construct data-driven surrogate models (SMs) using artificial neural networks in combination with concepts from many-body expansion.
The resulting SMs, developed using an active learning strategy known as query by committee, are validated through a variety of methods to ensure acceptable performance in estimating the hydrodynamic coefficients, (energy-related) objective function, and decision variables.
To rectify inherent errors in SMs, a hybrid optimization strategy is devised.
It involves solving an optimization problem with a genetic algorithm and SMs to generate a starting point that will be used with a gradient-based optimizer and MS.
The effectiveness and efficiency of the proposed approach are demonstrated by solving a series of optimization problems with increasing levels of complexity and integration for a $5$-WEC farm.
For a layout optimization study, the proposed framework offers a $91$-fold increase in computational efficiency compared to the direct usage of MS.
Previously unexplored investigations of much further complexity are also performed, leading to a concurrent geometry, farm- or device-level control, and layout optimization of heaving cylinder WEC devices in probabilistic irregular waves for a variety of coastal locations in the US.

The scalability of the method is evaluated by increasing the farm size to include $25$ WEC devices.
The results indicate promising directions toward a practical framework for integrated WEC farm design with more tractable computational demands.
\end{abstract}



\begin{keyword}

wave energy converter farms \sep surrogate modeling \sep control co-design \sep layout optimization \sep design coupling \sep machine learning




\end{keyword}

\end{frontmatter}

%% file: Nomenclature.tex


\nomenclature[A]{\(\textrm{WEC}\)}{wave energy converter}
\nomenclature[A]{\(\textrm{TRL}\)}{technology readiness level}
\nomenclature[A]{\(\textrm{CCD}\)}{control co-design}
\nomenclature[A]{\(\textrm{PTO}\)}{power take-off}
\nomenclature[A]{\(\textrm{SM}\)}{surrogate model/modeling}
\nomenclature[A]{\(\textrm{BEM}\)}{boundary element method}
\nomenclature[A]{\(\textrm{MS}\)}{multiple scattering}
\nomenclature[A]{\(\textrm{ANN}\)}{artificial neural networks}
\nomenclature[A]{\(\textrm{MBE}\)}{many-body expansion}
\nomenclature[A]{\(\textrm{PDF}\)}{probability distribution function}
\nomenclature[A]{\(\textrm{GA}\)}{genetic algorithm}
\nomenclature[A]{\(\textrm{CFD}\)}{computational fluid dynamics}
\nomenclature[A]{\(\textrm{QoI}\)}{
quantities of interest}
\nomenclature[A]{\(\textrm{LHS}\)}{
Latin hypercube sampling}
\nomenclature[A]{\(\textrm{QBC}\)}{
query by committee}
\nomenclature[A]{\(\textrm{MSE}\)}{
mean squared error}


\nomenclature[v, 01]{\(\AddedMass\)}{added mass coefficient matrix}
\nomenclature[v, 02]{\(\DampingCoeff\)}{damping coefficient matrix}
\nomenclature[v, 03]{\( \BPTO \)}{PTO damping coefficients}
\nomenclature[v, 05]{\( D_{wec} \)}{diameter of WEC device}
\nomenclature[v, 06]{\( \hat{\Force}_{\text{e}}\)}{excitation force}
\nomenclature[v, 07]{\( \hat{\Force}_{\text{FK}} \)}{Froude-Krylov force}
\nomenclature[v, 08]{\( \hat{\Force}_{\text{hs}}\)}{hydrostatic force}
\nomenclature[v, 09]{\( \hat{\Force}_{\text{pto}}\)}{power-take-off force}
\textbf{\nomenclature[v, 10]{\( \hat{\Force}_{\text{r}}\)}{radiation force}}
\nomenclature[v, 11]{\( \hat{\Force}_{\text{s}} \)}{scattering force}
\nomenclature[v, 12]{\( G \)}{hydrostatic coefficient}
\nomenclature[v, 13]{\( g \)}{gravitational acceleration}
\nomenclature[v, 14]{\( \mathbf{H} \)}{transfer function matrix}
\nomenclature[v, 15]{\( H_i \)}{wave amplitude}
\nomenclature[v, 16]{\( H_s \)}{significant wave height}
\nomenclature[v, 17]{\( h \)}{water depth}
\nomenclature[v, 18]{\( \KPTO \)}{PTO stiffness}
\nomenclature[v, 19]{\( k \)}{wave number}
\nomenclature[v, 20]{\( l_{pq} \)}{relative distance between $p$th and $q$th WEC}
\nomenclature[v, 21]{\( \Mass \)}{diagonal mass matrix}
\nomenclature[v, 22]{\( p_a \)}{average power}
\nomenclature[v, 23]{\( \mathbf{p}_{i} \)}{mechanical power matrix}
\nomenclature[v, 24]{\( p_v \)}{power per unit volume objective function}
\nomenclature[v, 25]{\( R_{wec} \)}{radius of the WEC device}
\nomenclature[v, 26]{\( RD_{wec} \)}{slenderness ratio}
\nomenclature[v, 27]{\( S \)}{cross-sectional area at undisturbed sea level}
\nomenclature[v, 28]{\( S_{JS}\)}{sea state described by JONSWAP}
\nomenclature[v, 29]{\( s_d \)}{safety distance}
\nomenclature[v, 30]{\( T_p \)}{wave period}
\nomenclature[v, 31]{\( \mathbf{w} \)}{dimensional layout matrix}
\nomenclature[v, 32]{\( x_p, y_p \)}{position of $p$th WEC}

\nomenclature[v, 35]{\( \beta_w \)}{angle of wave direction}
\nomenclature[v, 37]{\( \Delta\psi \)}{additive hydrodynamic interaction effect}
\nomenclature[v, 38]{\( \eta \)}{irregular incident wave field}
\nomenclature[v, 39]{\( \eta_{oa} \)}{operational availability}
\nomenclature[v, 40]{\( \eta_{pcc} \)}{efficiency of the power conversion chain}
\nomenclature[v, 41]{\( \eta_t \)}{transmission efficiency}
\nomenclature[v, 42]{\( \theta_{pq} \)}{relative angle between $p$th and $q$th WEC with respect to positive direction of x-axis}
\nomenclature[v, 43]{\( \hat{\bm{\xi}} \)}{displacement vector}
\nomenclature[v, 44]{\( \rho \)}{density of water}
\nomenclature[v, 45]{\( \Phi \)}{potential associated with waves}
\nomenclature[v, 46]{\( \psi \)}{hydrodynamic interaction effect}
\nomenclature[v, 47]{\( \omega \)}{angular frequency}
